\documentclass[twocolumn,prb,reprint,superscriptaddress,floatfix,amsmath,amssymb,aps]{revtex4-2}
\pdfoutput=1
\usepackage{array}

\usepackage[utf8]{inputenc}
\usepackage[caption=false,subrefformat=parens,labelformat=parens]{subfig}
\usepackage{graphicx}
\usepackage{float}
\usepackage{amsmath}
\usepackage{dcolumn}
\usepackage{bm}
\usepackage{xcolor}
\usepackage{xfrac}
\usepackage[colorlinks=true, allcolors=blue]{hyperref}
\usepackage{chemformula}
\usepackage{siunitx}
\usepackage{booktabs}

\begin{document}
\title{Uniaxial ferromagnetism in the kagome metal TbV${_6}$Sn${_6}$}

\author{Elliott Rosenberg}
\affiliation{Department of Physics, University of Washington, Seattle, Washington, 98195, USA}

\author{Jonathan M. DeStefano}
\affiliation{Department of Physics, University of Washington, Seattle, Washington, 98195, USA}

\author{Yucheng Guo}
\affiliation{%
 Department of Physics and Astronomy, Rice University, Houston, 77005 Texas, USA\\
}%

\author{Ji Seop Oh}
\affiliation{%
 Department of Physics and Astronomy, Rice University, Houston, 77005 Texas, USA\\
}%
\affiliation{%
 Department of Physics, University of California, Berkeley, 94720 California, USA\\
}%

\author{Makoto Hashimoto}
\affiliation{%
Stanford Synchrotron Radiation Lightsource, SLAC National Acelerator Laboratory, Menlo Park, 94025 California, USA
}%

\author{Donghui Lu}
\affiliation{%
Stanford Synchrotron Radiation Lightsource, SLAC National Acelerator Laboratory, Menlo Park, 94025 California, USA
}%

\author{Robert J. Birgeneau}
\affiliation{%
 Department of Physics, University of California, Berkeley, 94720 California, USA\\
}%

\author{Yongbin Lee}
\affiliation{Ames Laboratory, U.S. Department of Energy, Ames, Iowa 50011}

\author{Liqin Ke}
\affiliation{Ames Laboratory, U.S. Department of Energy, Ames, Iowa 50011}

\author{Ming Yi}
\affiliation{%
 Department of Physics and Astronomy, Rice University, Houston, 77005 Texas, USA\\
}%

\author{Jiun-Haw Chu}
\affiliation{Department of Physics, University of Washington, Seattle, Washington, 98195, USA}

\date{\today}

\begin{abstract}
The synthesis and characterization of the vanadium-based kagome metal \ch{TbV6Sn6} is presented. X-ray measurements confirm this material forms with the same crystal structure type as the recently investigated kagome metals GdV$_6$Sn$_6$ and YV$_6$Sn$_6$, with space group symmetry P6/mmm. A signature of a phase transition at 4.1K is observed in heat capacity, resistivity, and magnetic susceptibility measurements, and both resistivity and magnetization measurements exhibit hysteresis in magnetic field. Furthermore, a strikingly large anisotropy in the magnetic susceptibility was observed, with the c-axis susceptibility nearly 100 times the ab plane susceptibility at 2K. This is highly suggestive of uniaxial ferromagnetism, and the large size of 9.4$\mu_b$/f.u. indicates the Tb$^{3+}$ $4f$ electronic moments cooperatively align perpendicular to the V kagome lattice plane. The entropy at the phase transition is nearly Rln(2), indicating that the CEF ground state of the Tb$^{3+}$ ion is a doublet, and therefore the sublattice of $4f$ electrons in this material can be shown to map at low temperatures to the Ising model in a D$_{6h}$ symmetry environment. Hall measurements at temperatures from 300K to 1.7K can be described by two-band carrier transport at temperatures below around 150K, with a large increase in both hole and electron mobilities, similar to YV$_6$Sn$_6$, and an anomalous Hall effect is seen below the ordering temperature. Angle-resolved photoemission measurements above the magnetic ordering temperature reveal typical kagome dispersions. Our study presents \ch{TbV6Sn6} as an ideal system to study the interplay between Ising ferromagnetism and non-trivial electronic states emerging from a kagome lattice.
\end{abstract}

\maketitle

\section{Introduction}
Recently, metals containing kagome lattices have gained attention due to the appearance of Dirac points, van Hove singularities and geometrically driven flat bands in their resulting band structures~\cite{Ye2018,Kang2020,CoSn_comin, giantSOC_Hasan, negativeflatband_hasan,Ghimire2020,Ortiz2019,Ortiz2020}. Many of these materials are candidates to exhibit correlation driven topological magnetism. In particular, members of the RT$_6$X$_6$ subset of these materials form with kagome layers comprised solely of transition metal sites (T), potentially enabling the study of idealized pristine kagome lattices. These materials host an interplay between the nontrivial topological band structure effects that arise from the kagome sublattice and magnetism originating from potentially large $4f$ moments in the rare-earth (R) site and $3d$ moments in the transition metal sites. This includes the T=Mn set of recently investigated materials, in which TbMn$_6$Sn$_6$ exhibited robust Chern topological magnetism~\cite{TMS_hasan}, and the other rare earth substitutions have shown similar effects including a large anomalous Hall effect (AHE)~\cite{RMS_engineering_Jia,LMS_anomhall-FM_felser}. A similar set of materials, but with non-magnetic V substituted for Mn in the intermetallic T site, might very well promise similar physics with a cleaner separation of magnetic and electronic subsystems. Recently GdV$_6$Sn$_6$ and YV$_6$Sn$_6$ have been experimentally investigated~\cite{YVSGVS_wilson,Ishikawa2021}, with the Gd member undergoing ferromagnetic order at 5K, and a calculated band structure suggestive of hosting non-trivial topological phases. In addition, ScV$_6$Sn$_6$ was found to undergo a charge density wave phase transition reminiscent of the CsV$_3$Sb$_5$ system \cite{Arachchige2022}. Angle resolved photoemission (ARPES) experiments on \ch{GdV6Sn6} and \ch{HoV6Sn6} show the characteristic Dirac cone, saddle point, and flat bands of a kagome lattice from the purely vanadium kagome layer~\cite{GVSHVS_arpes_He}. This implies that RV$_6$Sn$_6$ presents an ideal platform to study kagome physics.

It has been recently noted that the kagome materials can be manipulated via different rare-earth substitutions to provide different lattice spacings as well as tune $4f$ interactions with the transition metal kagome network. Notably, anomalous Hall contributions to the Hall resistivity appear in some but not all of these materials~\cite{LMS_anomhall-FM_felser, RMS_anomhall-FIM_qi}, indicating the sensitive nature of the interplay between $4f$ magnetism and the geometrically driven topological band structure effects from the kagome lattice. In particular, the out-of-plane uniaxial ferrimagnetism in \ch{TbMn6Sn6}~\cite{RMS_many_gschneidner} is critical for realizing the spinless Haldane model that generates the Chern gapped Dirac fermions. Such a magnetic state has not been observed in the RV$_6$Sn$_6$ materials discovered so far. This motivates looking at the series of vanadium based kagome lattices, and we pursue this by synthesizing and characterizing TbV$_6$Sn$_6$ which exhibits out-of-plane uniaxial ferromagnetism at 4.1K.

\begin{figure*}
    \centering
    \includegraphics[width=0.8\textwidth]{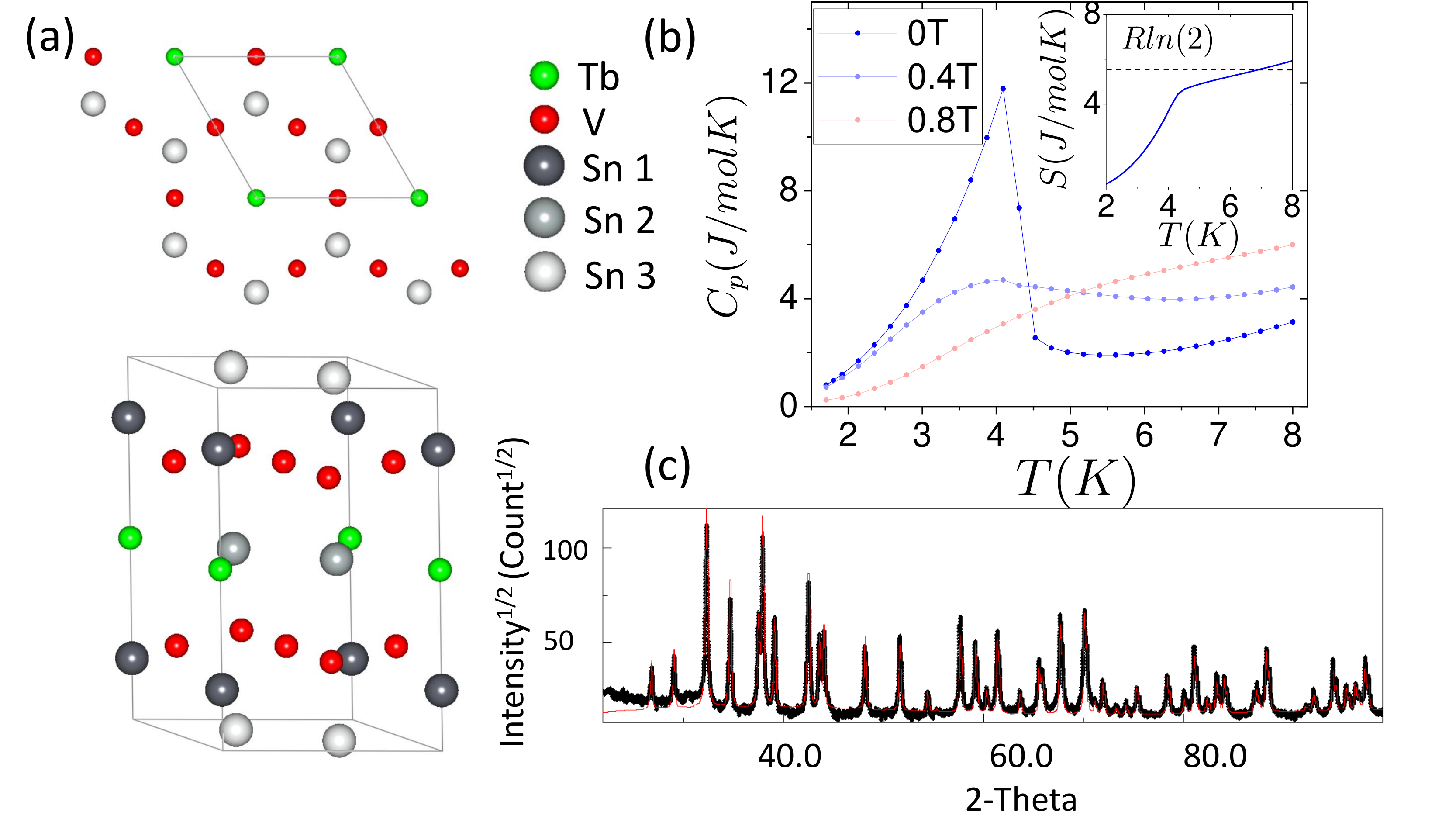}
    \caption{(a) The crystal structure of TbV$_6$Sn$_6$, which is of the MgFe$_6$Ge$_6$ crystal structure type with P6/mmm symmetry, with a view along the z-axis at the top (Tb sites are shown at the corners of the unit cell as opposed to Sn1 sites for clarity). (b) The heat capacity measurements at various magnetic fields applied along the c-axis from 0T to 0.8T. For the heat capacity at 0T a sharp mean-field like step is observed at 4.1K. The inset depicts the entropy (integrated from C$_p$/T, which was interpolated assuming it is a smooth monotonic function of temperature with value 0 at 0K). It approaches near the value Rln(2) at the phase transition. (c) The X-ray diffraction spectrum obtained by powdered samples, with the refined structure fit shown in red. }
    \label{fig:Cp}
\end{figure*}
\section{Methods}
Single crystals of TbV$_6$Sn$_6$ were synthesized via the self-flux method, with Sn used as the excess flux. Mixtures of Tb (99.999\%) pieces, V pieces (99.7\%) and Sn shot (99.999\%) were loaded into a fritted alumina crucible (CCS)\cite{Canfield2016}  with atomic ratios 4.5:27:95.5, then vacuum sealed in quartz tubes. These were heated up to 1200C, held there for 12 hours, then slowly cooled to 600C in 200 hours where the growths were decanted in a centrifuge to separate the excess flux from the crystals.

Powder X-ray diffraction measurements were performed on crushed single crystals using the Rigaku MiniFlex 600 system, with a Cu source and Hy-Pix 400MF 2D-detector. The lattice parameters and crystal structure type were determined via structural refinements using the MAUD software package \cite{MAUD}.

Heat capacity measurements were carried out using the Heat Capacity option of the Quantum Design Dynacool Physical Property Measurement System (PPMS). A 1.6mg plate-like single crystal was thermally anchored to the platform using N-grease.

Magnetization measurements were performed via the Vibrating Sample Magnetometer (VSM) option of the PPMS. For field in-plane measurements the sample was glued to a quartz paddle using Two-Part Epoxy (to prevent the sample from being torn off at high magnetic fields due to the torque from magnetic anisotropy). For field out of plane measurements samples were secured using G.E. Varnish on quartz pieces which were held by a brass sample holder. Multiple measurements on different samples were done to confirm the magnetic anisotropy of TbV$_6$Sn$_6$, although the data in Figure \ref{fig:M} is from the same sample for both in-plane and out-of-plane measurements to minimize relative errors from weighing.    

Transport measurements were performed on samples that were polished and cut by a wire saw to be of bar shape with dimensions roughly 1mm x 0.4mm x 0.05mm. HCl acid (6M) was used to remove excess Sn flux off the surface of the crystals. Silver paste and gold wires were used to make 4 point and 5 point (Hall pattern) measurements. These measurements were performed in a Dynacool PPMS in temperatures from 300-1.7K and in magnetic fields up to 14T. 

Angle-resolved photoemission (ARPES) measurements were performed using a DA30L analyzer at Beamline 5-2 of the Stanford Synchrotron Radiation Lightsource (SSRL) with an energy resolution of 15 meV and angular resolution of 0.1$^{\circ}$.  The samples were cleaved \textit{in-situ} and measured with the base pressure below 3×10$^{-11}$ torr at 15K.

To compare with band structures measured in ARPES, we calculated and projected bands onto the surface Brillouin zone (BZ) by integrating the ${\bf k}$-dependent spectral function $\delta[\omega-E_i({\bf k}_{\parallel}, k_z)]$ over $k_z$.
This is carried out using an in-house \textit{ab initio} tight-binding (TB) code~\cite{ke2019prb}.
Details of band structure and magnetocrystalline calculations can be found elsewhere~\cite{lee2022arxiv}.

\section{Results and Discussion}

\begin{table}[]

\begin{tabular}{c c c c c}
\toprule
atom(site) & x     & y     & z     & Occupancy \\ \midrule
Tb         & 1     & 1     & 0.5   & 1         \\
V          & 0.5   & 0.5     & 0.7515 & 1         \\
Sn1        & 1     & 1 & 0.8317     & 1         \\
Sn2        & 0.3333 & 0.6667 & 0.5   & 1         \\
Sn3        & 0.3333     & 0.6667     & 1 & 1         \\ \bottomrule
\end{tabular}
\caption{Microstructure parameters determined for the TbV$_6$Sn$_6$ unit cell in the P6/mmm hexagonal space group symmetry. The lattice constants were determined to be  $a= 5.531 \pm 0.001 \AA $ and $c= 9.199 \pm 0.002 \AA$. }
\label{tab1}
\end{table}

\subsection{X-ray}
The crystal structure and lattice constants were obtained by the refinement of powder X-ray diffraction measurements of crushed single crystals of TbV$_{6}$Sn$_{6}$. Refinement solutions were determined using the MAUD software package \cite{MAUD}. Impurity phases including Sn residue were determined to be less than 1\% of the total material measured. Cell refinement in the hexagonal space group P6/mmm determined the lattice constants to be $a= 5.531 \pm 0.001 \AA $ and $c= 9.199 \pm 0.002 \AA$. The crystal structure of TbV$_{6}$Sn$_{6}$ is shown in Figure \ref{fig:Cp}(a). The microstructure parameters derived from the refinement are listed in Table \ref{tab1}. The derived crystal structure is consistent with previous measurements of the rare-earth substituted GdV$_6$Sn$_6$ and YV$_6$Sn$_6$~\cite{YVSGVS_wilson}.

\subsection{Heat capacity}
The heat capacity measurements are summarized in Figure \ref{fig:Cp}(b). There is a sharp mean-field like step observed at 4.1K. This step is broadened considerably by the application of relatively small magnetic fields, until the transition is no longer observable above around 0.5T. This, along with magnetic measurements shown in Figure \ref{fig:M}, is suggestive that the conjugate field to the phase transition is a magnetic field along the c-axis. The entropy (integrated from C$_p$/T, which was interpolated assuming it is a smooth monotonic function of temperature with value 0 at 0K) approaches a value near Rln(2) at the phase transition. This suggests the phase transition is Ising-like in that it involves the splitting of a degenerate magnetic doublet. The shape of the heat capacity below the phase transition also strongly resembles the heat capacity signature of an canonical spin-1/2 Ising phase transition.

\subsection{Magnetic susceptibility and magnetization versus field (MvH)}
\begin{figure*}
    \centering
    \includegraphics[width=0.9\textwidth]{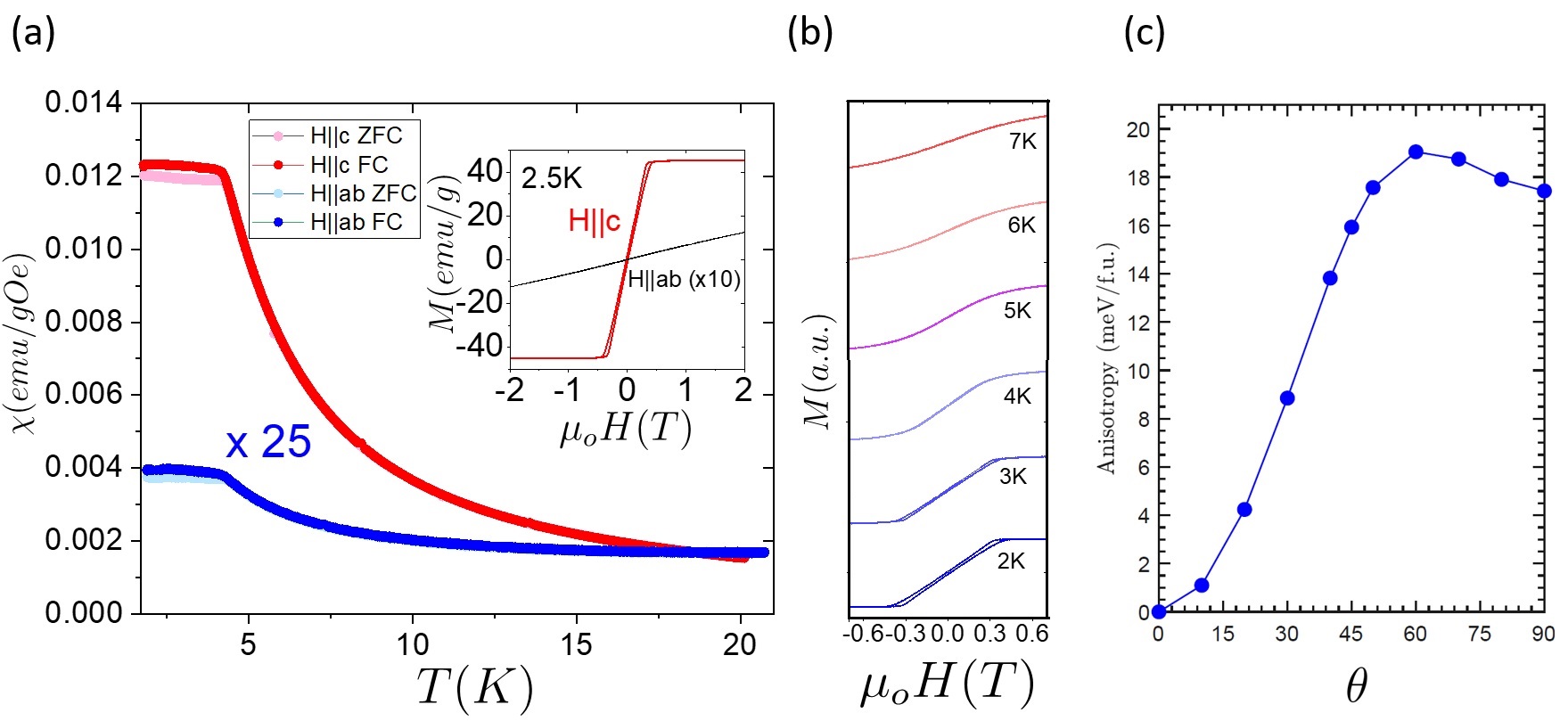}
    \caption{(a) Magnetic susceptibility measured at 500 Oe, with the red curves indicating the magnetic field is along the c-axis and the blue curves (multiplied by 25) with the field along the ab-plane. ZFC and FC measurements were performed, and hysteretic behavior indicative of domains is present for both orientations. The inset contains moment versus field data at 2.5K with the field along the c-axis in red and the field along the ab-plane in black (multiplied by 10). (b) The magnetic moment in arbitrary units versus magnetic field applied along the c-axis at temperatures from 2 to 7K, offset for clarity. Hysteresis is observed at temperatures below the phase transition at 4.1K, but not above.}
    \label{fig:M}
\end{figure*}

Magnetic susceptibility measurements performed at 500 Oe are depicted in Figure \ref{fig:M}(a), with the red curves indicating the magnetic field is along the c-axis and the blue curves (multiplied by 25) with the field along ab-plane. ZFC and FC measurements were performed, and hysteretic behavior indicative of domains is present for both orientations. However there is a striking magnetic anisotropy evident in this material, as the magnetic susceptibility along the c-axis is nearly 100 times the magnetic susceptibility along the ab-plane in the proximity of the phase transition. Even up to 200K the c-axis susceptibility remains more than twice as large as the in-plane susceptibility. MvH measurements are also displayed in Figure \ref{fig:M}(b) with H oriented along the c-axis from 2K to 7K, and there are clear signs of hysteresis for temperatures below 4K, but not above. In-plane MvH measurements performed up to 14T (shown up to 2T in the inset to Figure \ref{fig:M}(a)) can be extrapolated to determine the magnetic anisotropy energy of this compound at nearly 330K/f.u., larger than that of canonical uniaxial ferromagnets like LiHoF$_4$~\cite{LiHoF4_walker}.

The magnetic anisotropy energy was also calculated from first principles, and the results as a function of angle $\theta$ from the c-axis is shown in Figure \ref{fig:M}(c). The difference in the anisotropy energy between 0$^{\circ}$ and 90$^{\circ}$ was calculated to be roughly 17.5meV/f.u. (203K/f.u.), which is consistent with the large value found from MvH measurements.

\subsection{Resistivity and magnetotransport}

\begin{figure*}
    \centering
    \includegraphics[width=0.8\textwidth]{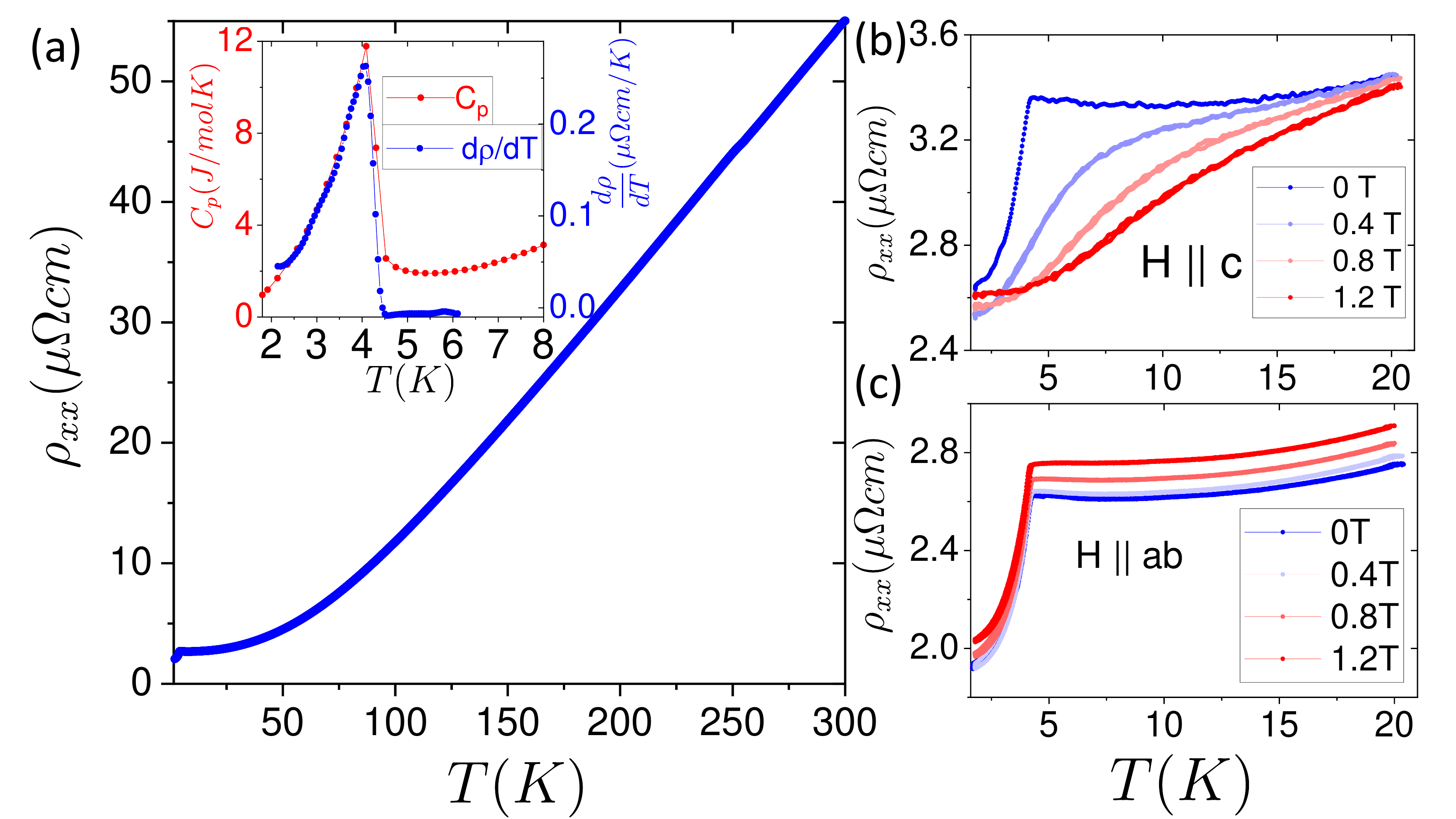}
    \caption{(a) In-plane resistivity versus temperature from 2K to 300K. (b) Resistivity versus temperature data with magnetic fields applied along the c-axis, and similarly to the heat capacity the signature of the phase transition is broadened considerably above 0.5T. (c) Resistivity versus temperature curves with magnetic field applied along the ab-plane, and the application of this field does not shift or broaden the phase transition noticeably.}
    \label{fig:res}
\end{figure*}

The in-plane resistivity versus temperature is shown in Figure \ref{fig:res}(a). There is a clear and pronounced kink at 4.1K, shown more clearly in Figure \ref{fig:res}(b) at 0T. In the inset of Figure \ref{fig:res}(a), the heat capacity and d$\rho$/dT are shown, and display a similar temperature dependence in the proximity of the ferromagnetic phase transition, indicating the classic Fisher-Langer argument is relevant for this compound. This implies the magnetic fluctuations that result from the phase transition play a similarly dominant role in determining both the energy of the system and the scattering rate of the longitudinal resistivity at low temperatures. Figure \ref{fig:res}(b) displays resistivity versus temperature data with magnetic fields applied along the c-axis, and similarly to the heat capacity the signature of the phase transition is broadened considerably above 0.5T. Figure \ref{fig:res}(c) displays resistivity versus temperature curves with magnetic field applied along the ab-plane, and the application of fields up to 1.2T does not shift or broaden the phase transition noticeably.

\subsection{Hall measurements}
\begin{figure}
    \centering
    \includegraphics[width=0.5\textwidth]{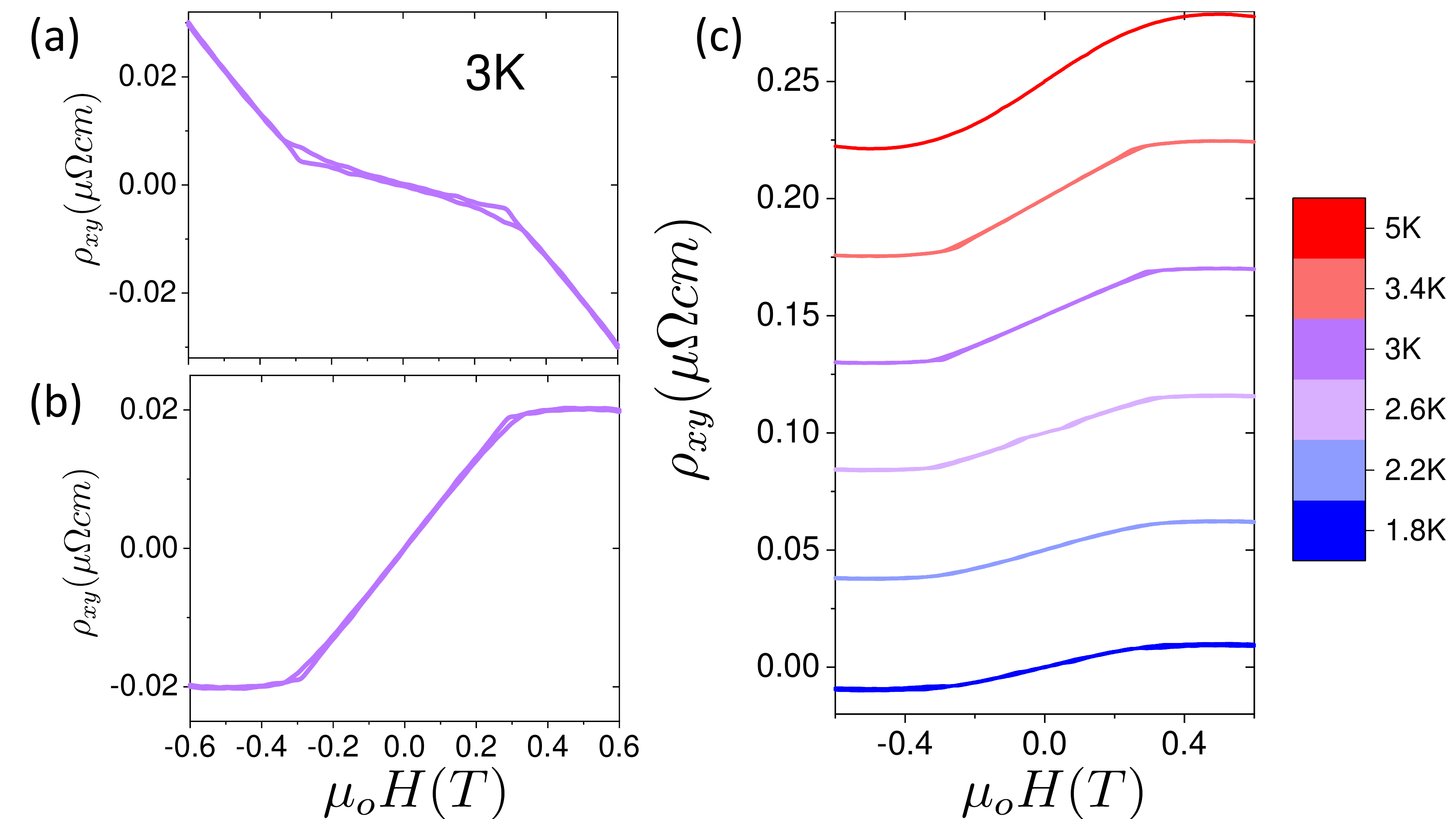}
    \caption{(a) Anti-symmetrized $\rho_{xy}$ data versus field at low fields at 3K. There is a noticeable hysteresis in the data, indicative of an anomalous Hall contribution. (b) Same as (a) but subtracted by a high-field linear fit. (c) Similar data (offset for clarity) with subtracted linear backgrounds at temperatures from 1.8 to 5K, suggestive that an anomalous Hall contribution exists at temperatures below and above the ferromagnetic phase transition.}
    \label{fig:AHE}
\end{figure}

\begin{figure*}
    \centering
    \includegraphics[width=0.8\textwidth]{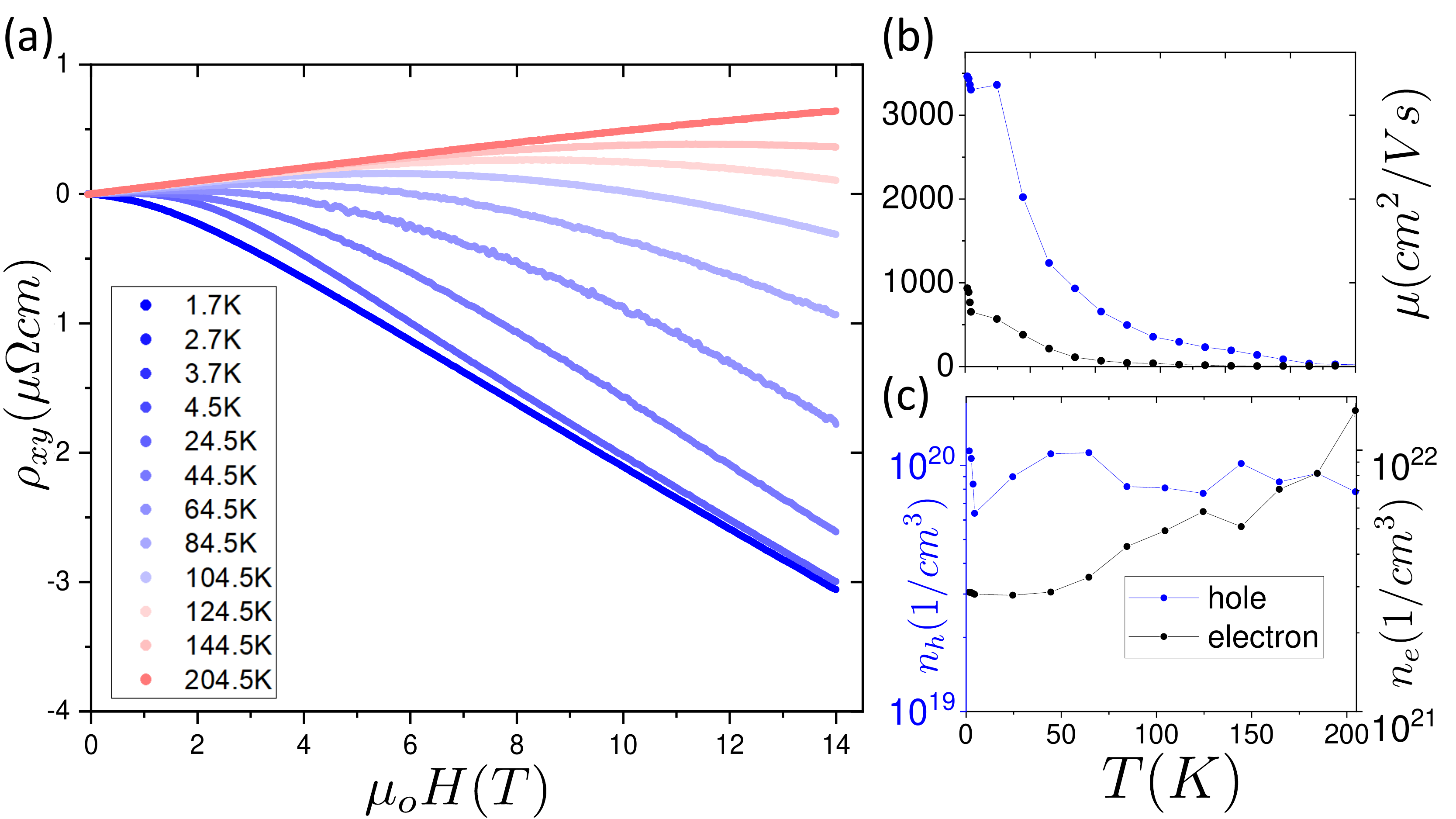}
    \caption{(a) Hall measurements from 200K to 1.7K with fields sweeping from 0T to 14T, then down to -14T and back to 0T. Only the anti-symmetrized data at positive fields is shown. (b) Mobilities and (c) carrier densities extracted from two-band fitting of Hall measurements. Both the electron and hole mobilities increase substantially as the temperature is decreased, consistent with previous measurements of YV$_6$Sn$_6$. }
    \label{fig:Hall}
\end{figure*}

Hall measurements were performed from 300K to 1.7K with fields being swept in both directions. We first focus on low field data near the transition temperature where an AHE emerges. As shown in Figure \ref{fig:AHE}(a), there is hysteresis in $\rho_{xy}$ below 4K at field scales similar to those observed to be relevant to the hysteresis of the magnetization. Subtracting the linear background above the saturation field reveals the close resemblance between $\Delta\rho_{xy}$ (Figure \ref{fig:AHE}(b,c)) and magnetization (Figure \ref{fig:M}), unambiguously establishing the existence of an AHE. We next turn to the high field data, shown in Figure \ref{fig:Hall}, where $\rho_{xy}$ is dominated by the ordinary Hall effect. As the magnetic field increases well beyond the saturation field (0.5T), the ordinary Hall resistivity exhibits a non-linear magnetic field dependence at temperatures up to around 150K, indicating two-band behavior. Following the procedure described in \cite{Hallpaper}, two-band fitting was performed to extract the hole and electron mobilities and carrier densities, shown in the right panels of Figure \ref{fig:Hall}. Although the fitting is not quantitatively precise, especially at temperatures in which the Hall resistivity can be described effectively by a one-band model, we can still draw robust conclusions that the hole and electron mobilities increase significantly as temperature is decreased, and that the electron carrier density is larger than the hole carrier density at all temperatures, which leads to a positive Hall coefficient at temperatures where effective one band Hall transport is observed. 

\subsection{ARPES measurements}

\begin{figure*}
    \centering
    \includegraphics[width=\textwidth]{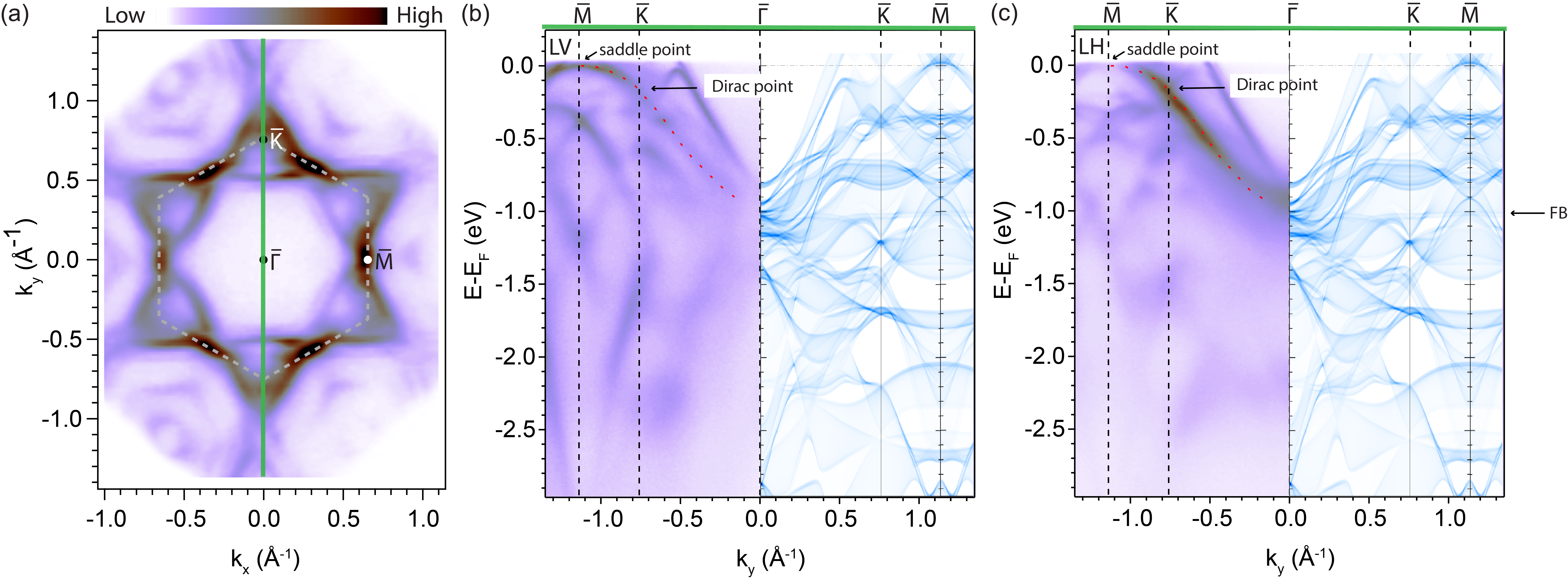}
    \caption{Electronic structure of the kagome termination of TbV$_6$Sn$_6$. (a) Fermi surface of TbV$_6$Sn$_6$ measured with linear horizontal (LH) polarization. White dashed line denotes the first Brillouin zone. (b) $\bar{\Gamma}$-$\bar{K}$-$\bar{M}$ cut measured with linear vertical (LV) polarization. DFT calculated dispersions along the same cut integrated along $k_z$ are shown for comparison. (c) Same as (b) but measured with LH polarization. Dirac point and saddle point are marked. All measurements were performed with 83 eV photons at 15K.
}
    \label{fig:arpes}
\end{figure*}

To investigate the electronic structure, we measured the normal state electronic structure of TbV$_6$Sn$_6$ at 15 K using ARPES (Figure \ref{fig:arpes}). We found two types of terminations, distinguished by the different valence band structures as well as core level spectroscopy. Consistent with previous work on Ho/GdV$_6$Sn$_6$~\cite{GVSHVS_arpes_He}, the two terminations correspond to the kagome termination and the Sn termination. In Figure \ref{fig:arpes} we present the results from the kagome termination. Typical of vanadium kagome systems, the Fermi surface exhibits triangular pockets centered at the K points of the Brillouin zone (BZ) with tips of the triangles meeting at the M points. Band dispersions along high symmetry directions are also shown, taken with different polarizations (Figure \ref{fig:arpes} (b)-(c)). For comparison we also show the DFT calculated bulk bands integrated along k$_z$. There is overall good agreement between calculated and measured dispersions. In particular, we identify the presence of a Dirac crossing near -0.2 eV at the K point that disperses up to a saddle point near the Fermi level at the M point, which is also captured in the DFT calculations. The overall electronic structure of the normal state of TbV$_6$Sn$_6$ resembles that of Ho/GdV$_6$Sn$_6$~\cite{GVSHVS_arpes_He}. We also note that the kagome flat bands according to DFT appear near -1eV. We do not observe strong intensity of the flat bands in this energy range, possibly due to the strong $k_z$ dispersion of the kagome flat bands resulting from finite interlayer interactions, which coupled with the $k_z$ broadening effect due to the poor out-of-plane resolution of the photoemission process makes the dispersive flat bands harder to resolve. 

\section{Discussion}

 YV$_6$Sn$_6$, which is nearly identical to TbV$_6$Sn$_6$ except the rare-earth site has been replaced with an ion with no $4f$ electrons, has been measured recently and found to have no magnetic order or evident magnetic anisotropy down to 1.7K \cite{YVSGVS_wilson}. Thus the cause of the ferromagnetic phase transition in TbV$_6$Sn$_6$ is from the cooperative alignment of the $4f$ electrons of the Tb$^{3+}$ ion. The magnetic anisotropy seen in this material can be understood as a consequence of the relevant energy scales for the $4f$ sublattice.
 
 The finer details of $4f$ energy scales have been explained elsewhere~\cite{quadrupoleinteractions_morin, transising_maharaj}, but here is a brief summary. The hierarchy of energy scales starts with Coulomb repulsion (which manifests in the Hunds' rules which determine the filling of the $4f$ orbitals.) Spin-orbit coupling determines the total angular momentum number J=L+S to be the good quantum number for the $4f$ electronic multiplet, and then the perturbative crystalline electric field (CEF) potential from the surrounding environment splits the 2J+1 degenerate states.  The multiplicity of the degeneracies of the CEF spectrum is determined by the point group symmetry surrounding the $4f$ ion. In the case of the Tb ion in a D$_{6h}$ point group symmetry, the 8 $4f$ electrons form a J=6 electronic multiplet, and the 13 degenerate states are split into 5 singlets and 4 doublets. However only the degeneracies can be known $\textit{a priori}$ (from the dimensions of the corresponding irreducible representations), but the eigenenergies cannot be determined without knowing microscopic details of the surrounding ligands.   

Group theory considerations (similar argument as seen in Supplemental Material of \cite{transising_maharaj}) also fix the constituent m$_j$ states of the different CEF states. In the case of the two different types (irreducible representations) of doublets allowed in D$_{6h}$ for J=6, the E$_{1g}$ doublet is composed of $m_j=\pm 4$ and $\pm 2$ states, and the E$_{2g}$ doublet is composed of $\pm 5$ and $\pm 1$ states. A CEF spectrum which would explain both the heat capacity and the extreme magnetic anisotropy, would be one in which the ground state is one of the aforementioned doublets, and that the next nearest excited state was $\Delta_{CEF}>100K$ above. Common energy scales for CEF splittings can be 100K or higher.  Because $\langle E_{ig}^n | J_x | E_{ig}^m \rangle =0$ for both doublets E$_{1g}$ and E$_{2g}$, (n and m here represent each doublet state), there would be no magnetic susceptibility in the ab-plane from the $4f$ ground state, and the susceptibility would only become more isotropic at temperatures at and above $\Delta_{CEF}$. This ground state could of course be polarized in the z-direction, and $4f$ sublattice sites could cooperatively align ferromagnetically along the c-axis, providing an entropy of Rln(2) at the phase transition. A more thorough calculation was performed to assess the magnetic energy of the 4$f$ electrons of Tb as a function of angle from the z-axis (depicted in Figure \ref{fig:M}(c)), explicitly considering the Hamiltonian arising from considering both spin-orbit coupling and crystal field terms that can be approximated from the atomic spacings.  

An intriguing question is whether the Ising ferromagnetism in TbV$_6$Sn$_6$ may lead to the formation of exotic phases, such as Chern gapped Dirac fermions in the bulk, or a quantum anomalous Hall surface state. Previous density functional theory calculations suggested that the electronic structures of several RV$_6$Sn$_6$ compounds can be characterized as that of a topological metal, with surface states originating from the non-trivial $\mathbb{Z}_2$ indices associated with the occupied bands \cite{YVSGVS_wilson}. Our ARPES measurements are consistent with others that have revealed bulk bands with Dirac dispersion near the K point of the Brillouin zone \cite{GVSHVS_arpes_He}. The observed anomalous Hall effect of TbV$_6$Sn$_6$ is a clear signature of the impact of magnetism on the itinerant electrons. Using the anomalous Hall resistivity and longitudinal resistivity, we calculated the value of anomalous Hall conductivity at 1.8K to be approximately 2000 $\Omega^{-1}\cdot cm^{-1}$. This value, together with the value of the conductivity, puts the anomalous Hall effect of TbV$_6$Sn$_6$ at the boundary between being dominated by skew-scattering or arising from an intrinsic effect \cite{AHEpaper}. In principle, the intrinsic contribution of AHE can be extracted by a scaling analysis if it can be obtained in a parameter range where the resistivity varies by orders of magnitude. However, due to the low Curie temperature, the longitudinal resistivity changes very little in the temperature range where we observed the AHE. Future studies of AHE as a function of mean free path by systematically introducing disorder are needed to address this issue.

\section{Conclusions}
The synthesis of the vanadium-based kagome metal TbV$_6$Sn$_6$ single crystals is presented, and it was determined via powder X-ray diffraction to be of the same crystal structure type (MgFe$_6$Ge$_6$) and have the same space group symmetry (P6/mmm) as the recently discovered YV$_6$Sn$_6$ and GdV$_6$Sn$_6$. It was shown to have a ferromagnetic phase transition at 4.1K likely involving the cooperative alignment of the $4f$ electrons of the Tb$^{3+}$ ion. This compound displays a striking magnetic anisotropy, with magnetic susceptibility and MvH measurements confirming the c-axis is the magnetic easy axis and an extrapolated magnetic anisotropy energy of above 330K at 2.5K, making this material one of the ``most uniaxial" ferromagnets known at low temperatures. The heat capacity signature and entropy (extrapolated) at the phase transition provide evidence that it is an effective spin 1/2 Ising phase transition, possibly involving the spontaneous splitting of a $4f$ ground state doublet. The nature of this phase transition, and that of similar $4f$ magnetic phase transitions, which in general can be tuned towards a quantum phase transition via the application of effective transverse fields, inspires investigation into potential $4f$ quantum criticality in kagome metals.

\section{Ackowledgements}

This work is supported by the Air Force Office of Scientific Research under grant FA9550-21-1-0068, the David and Lucile Packard Foundation and the Gordon and Betty Moore Foundation’s EPiQS Initiative, grant no. GBMF6759 to J.-H.C. The work at Berkeley and LBNL was funded by the U.S. Department of Energy, Office of Science, Office of Basic Energy Sciences, Materials Sciences and Engineering Division under Contract No. DE-AC02-05-CH11231 (Quantum Materials program KC2202). The work at Rice was funded by the Robert A. Welch Foundation Grant No. C-2024 and the Gordon and Betty Moore Foundation's EPiQS Initiative through grant No. GBMF9470. Use of the Stanford Synchrotron Radiation Lightsource, SLAC National Accelerator Laboratory, is supported by the U.S. Department of Energy, Office of Science, Office of Basic Energy Sciences under Contract No. DE-AC02-76SF00515. LK and YL are supported by the U.S.~Department of Energy, Office of Science, Office of Basic Energy Sciences, Materials Sciences and Engineering Division, and Early Career Research Program.

\bibliography{main}

\begin{thebibliography}{26}%
\makeatletter
\providecommand \@ifxundefined [1]{%
 \@ifx{#1\undefined}
}%
\providecommand \@ifnum [1]{%
 \ifnum #1\expandafter \@firstoftwo
 \else \expandafter \@secondoftwo
 \fi
}%
\providecommand \@ifx [1]{%
 \ifx #1\expandafter \@firstoftwo
 \else \expandafter \@secondoftwo
 \fi
}%
\providecommand \natexlab [1]{#1}%
\providecommand \enquote  [1]{``#1''}%
\providecommand \bibnamefont  [1]{#1}%
\providecommand \bibfnamefont [1]{#1}%
\providecommand \citenamefont [1]{#1}%
\providecommand \href@noop [0]{\@secondoftwo}%
\providecommand \href [0]{\begingroup \@sanitize@url \@href}%
\providecommand \@href[1]{\@@startlink{#1}\@@href}%
\providecommand \@@href[1]{\endgroup#1\@@endlink}%
\providecommand \@sanitize@url [0]{\catcode `\\12\catcode `\$12\catcode
  `\&12\catcode `\#12\catcode `\^12\catcode `\_12\catcode `\%12\relax}%
\providecommand \@@startlink[1]{}%
\providecommand \@@endlink[0]{}%
\providecommand \url  [0]{\begingroup\@sanitize@url \@url }%
\providecommand \@url [1]{\endgroup\@href {#1}{\urlprefix }}%
\providecommand \urlprefix  [0]{URL }%
\providecommand \Eprint [0]{\href }%
\providecommand \doibase [0]{https://doi.org/}%
\providecommand \selectlanguage [0]{\@gobble}%
\providecommand \bibinfo  [0]{\@secondoftwo}%
\providecommand \bibfield  [0]{\@secondoftwo}%
\providecommand \translation [1]{[#1]}%
\providecommand \BibitemOpen [0]{}%
\providecommand \bibitemStop [0]{}%
\providecommand \bibitemNoStop [0]{.\EOS\space}%
\providecommand \EOS [0]{\spacefactor3000\relax}%
\providecommand \BibitemShut  [1]{\csname bibitem#1\endcsname}%
\let\auto@bib@innerbib\@empty
\bibitem [{\citenamefont {Ye}\ \emph {et~al.}(2018)\citenamefont {Ye},
  \citenamefont {Kang}, \citenamefont {Liu}, \citenamefont {von Cube},
  \citenamefont {Wicker}, \citenamefont {Suzuki}, \citenamefont {Jozwiak},
  \citenamefont {Bostwick}, \citenamefont {Rotenberg}, \citenamefont {Bell},
  \citenamefont {Fu}, \citenamefont {Comin},\ and\ \citenamefont
  {Checkelsky}}]{Ye2018}%
  \BibitemOpen
  \bibfield  {author} {\bibinfo {author} {\bibfnamefont {L.}~\bibnamefont
  {Ye}}, \bibinfo {author} {\bibfnamefont {M.}~\bibnamefont {Kang}}, \bibinfo
  {author} {\bibfnamefont {J.}~\bibnamefont {Liu}}, \bibinfo {author}
  {\bibfnamefont {F.}~\bibnamefont {von Cube}}, \bibinfo {author}
  {\bibfnamefont {C.~R.}\ \bibnamefont {Wicker}}, \bibinfo {author}
  {\bibfnamefont {T.}~\bibnamefont {Suzuki}}, \bibinfo {author} {\bibfnamefont
  {C.}~\bibnamefont {Jozwiak}}, \bibinfo {author} {\bibfnamefont
  {A.}~\bibnamefont {Bostwick}}, \bibinfo {author} {\bibfnamefont
  {E.}~\bibnamefont {Rotenberg}}, \bibinfo {author} {\bibfnamefont {D.~C.}\
  \bibnamefont {Bell}}, \bibinfo {author} {\bibfnamefont {L.}~\bibnamefont
  {Fu}}, \bibinfo {author} {\bibfnamefont {R.}~\bibnamefont {Comin}},\ and\
  \bibinfo {author} {\bibfnamefont {J.~G.}\ \bibnamefont {Checkelsky}},\
  }\bibfield  {title} {\bibinfo {title} {Massive dirac fermions in a
  ferromagnetic kagome metal},\ }\href {https://doi.org/10.1038/nature25987}
  {\bibfield  {journal} {\bibinfo  {journal} {Nature}\ }\textbf {\bibinfo
  {volume} {555}},\ \bibinfo {pages} {638} (\bibinfo {year}
  {2018})}\BibitemShut {NoStop}%
\bibitem [{\citenamefont {Kang}\ \emph
  {et~al.}(2020{\natexlab{a}})\citenamefont {Kang}, \citenamefont {Ye},
  \citenamefont {Fang}, \citenamefont {You}, \citenamefont {Levitan},
  \citenamefont {Han}, \citenamefont {Facio}, \citenamefont {Jozwiak},
  \citenamefont {Bostwick}, \citenamefont {Rotenberg}, \citenamefont {Chan},
  \citenamefont {McDonald}, \citenamefont {Graf}, \citenamefont {Kaznatcheev},
  \citenamefont {Vescovo}, \citenamefont {Bell}, \citenamefont {Kaxiras},
  \citenamefont {van~den Brink}, \citenamefont {Richter}, \citenamefont
  {Prasad~Ghimire}, \citenamefont {Checkelsky},\ and\ \citenamefont
  {Comin}}]{Kang2020}%
  \BibitemOpen
  \bibfield  {author} {\bibinfo {author} {\bibfnamefont {M.}~\bibnamefont
  {Kang}}, \bibinfo {author} {\bibfnamefont {L.}~\bibnamefont {Ye}}, \bibinfo
  {author} {\bibfnamefont {S.}~\bibnamefont {Fang}}, \bibinfo {author}
  {\bibfnamefont {J.-S.}\ \bibnamefont {You}}, \bibinfo {author} {\bibfnamefont
  {A.}~\bibnamefont {Levitan}}, \bibinfo {author} {\bibfnamefont
  {M.}~\bibnamefont {Han}}, \bibinfo {author} {\bibfnamefont {J.~I.}\
  \bibnamefont {Facio}}, \bibinfo {author} {\bibfnamefont {C.}~\bibnamefont
  {Jozwiak}}, \bibinfo {author} {\bibfnamefont {A.}~\bibnamefont {Bostwick}},
  \bibinfo {author} {\bibfnamefont {E.}~\bibnamefont {Rotenberg}}, \bibinfo
  {author} {\bibfnamefont {M.~K.}\ \bibnamefont {Chan}}, \bibinfo {author}
  {\bibfnamefont {R.~D.}\ \bibnamefont {McDonald}}, \bibinfo {author}
  {\bibfnamefont {D.}~\bibnamefont {Graf}}, \bibinfo {author} {\bibfnamefont
  {K.}~\bibnamefont {Kaznatcheev}}, \bibinfo {author} {\bibfnamefont
  {E.}~\bibnamefont {Vescovo}}, \bibinfo {author} {\bibfnamefont {D.~C.}\
  \bibnamefont {Bell}}, \bibinfo {author} {\bibfnamefont {E.}~\bibnamefont
  {Kaxiras}}, \bibinfo {author} {\bibfnamefont {J.}~\bibnamefont {van~den
  Brink}}, \bibinfo {author} {\bibfnamefont {M.}~\bibnamefont {Richter}},
  \bibinfo {author} {\bibfnamefont {M.}~\bibnamefont {Prasad~Ghimire}},
  \bibinfo {author} {\bibfnamefont {J.~G.}\ \bibnamefont {Checkelsky}},\ and\
  \bibinfo {author} {\bibfnamefont {R.}~\bibnamefont {Comin}},\ }\bibfield
  {title} {\bibinfo {title} {Dirac fermions and flat bands in the ideal kagome
  metal \ch{FeSn}},\ }\href {https://doi.org/10.1038/s41563-019-0531-0}
  {\bibfield  {journal} {\bibinfo  {journal} {Nature Materials}\ }\textbf
  {\bibinfo {volume} {19}},\ \bibinfo {pages} {163} (\bibinfo {year}
  {2020}{\natexlab{a}})}\BibitemShut {NoStop}%
\bibitem [{\citenamefont {Kang}\ \emph
  {et~al.}(2020{\natexlab{b}})\citenamefont {Kang}, \citenamefont {Fang},
  \citenamefont {Ye}, \citenamefont {Po}, \citenamefont {Denlinger},
  \citenamefont {Jozwiak}, \citenamefont {Bostwick}, \citenamefont {Rotenberg},
  \citenamefont {Kaxiras}, \citenamefont {Checkelsky},\ and\ \citenamefont
  {Comin}}]{CoSn_comin}%
  \BibitemOpen
  \bibfield  {author} {\bibinfo {author} {\bibfnamefont {M.}~\bibnamefont
  {Kang}}, \bibinfo {author} {\bibfnamefont {S.}~\bibnamefont {Fang}}, \bibinfo
  {author} {\bibfnamefont {L.}~\bibnamefont {Ye}}, \bibinfo {author}
  {\bibfnamefont {H.~C.}\ \bibnamefont {Po}}, \bibinfo {author} {\bibfnamefont
  {J.}~\bibnamefont {Denlinger}}, \bibinfo {author} {\bibfnamefont
  {C.}~\bibnamefont {Jozwiak}}, \bibinfo {author} {\bibfnamefont
  {A.}~\bibnamefont {Bostwick}}, \bibinfo {author} {\bibfnamefont
  {E.}~\bibnamefont {Rotenberg}}, \bibinfo {author} {\bibfnamefont
  {E.}~\bibnamefont {Kaxiras}}, \bibinfo {author} {\bibfnamefont {J.~G.}\
  \bibnamefont {Checkelsky}},\ and\ \bibinfo {author} {\bibfnamefont
  {R.}~\bibnamefont {Comin}},\ }\bibfield  {title} {\bibinfo {title}
  {Topological flat bands in frustrated kagome lattice {{CoSn}}},\ }\href
  {https://doi.org/10.1038/s41467-020-17465-1} {\bibfield  {journal} {\bibinfo
  {journal} {Nature Communications}\ }\textbf {\bibinfo {volume} {11}},\
  \bibinfo {pages} {4004} (\bibinfo {year} {2020}{\natexlab{b}})}\BibitemShut
  {NoStop}%
\bibitem [{\citenamefont {Yin}\ \emph {et~al.}(2018)\citenamefont {Yin},
  \citenamefont {Zhang}, \citenamefont {Li}, \citenamefont {Jiang},
  \citenamefont {Chang}, \citenamefont {Zhang}, \citenamefont {Lian},
  \citenamefont {Xiang}, \citenamefont {Belopolski}, \citenamefont {Zheng},
  \citenamefont {Cochran}, \citenamefont {Xu}, \citenamefont {Bian},
  \citenamefont {Liu}, \citenamefont {Chang}, \citenamefont {Lin},
  \citenamefont {Lu}, \citenamefont {Wang}, \citenamefont {Jia}, \citenamefont
  {Wang},\ and\ \citenamefont {Hasan}}]{giantSOC_Hasan}%
  \BibitemOpen
  \bibfield  {author} {\bibinfo {author} {\bibfnamefont {J.-X.}\ \bibnamefont
  {Yin}}, \bibinfo {author} {\bibfnamefont {S.~S.}\ \bibnamefont {Zhang}},
  \bibinfo {author} {\bibfnamefont {H.}~\bibnamefont {Li}}, \bibinfo {author}
  {\bibfnamefont {K.}~\bibnamefont {Jiang}}, \bibinfo {author} {\bibfnamefont
  {G.}~\bibnamefont {Chang}}, \bibinfo {author} {\bibfnamefont
  {B.}~\bibnamefont {Zhang}}, \bibinfo {author} {\bibfnamefont
  {B.}~\bibnamefont {Lian}}, \bibinfo {author} {\bibfnamefont {C.}~\bibnamefont
  {Xiang}}, \bibinfo {author} {\bibfnamefont {I.}~\bibnamefont {Belopolski}},
  \bibinfo {author} {\bibfnamefont {H.}~\bibnamefont {Zheng}}, \bibinfo
  {author} {\bibfnamefont {T.~A.}\ \bibnamefont {Cochran}}, \bibinfo {author}
  {\bibfnamefont {S.-Y.}\ \bibnamefont {Xu}}, \bibinfo {author} {\bibfnamefont
  {G.}~\bibnamefont {Bian}}, \bibinfo {author} {\bibfnamefont {K.}~\bibnamefont
  {Liu}}, \bibinfo {author} {\bibfnamefont {T.-R.}\ \bibnamefont {Chang}},
  \bibinfo {author} {\bibfnamefont {H.}~\bibnamefont {Lin}}, \bibinfo {author}
  {\bibfnamefont {Z.-Y.}\ \bibnamefont {Lu}}, \bibinfo {author} {\bibfnamefont
  {Z.}~\bibnamefont {Wang}}, \bibinfo {author} {\bibfnamefont {S.}~\bibnamefont
  {Jia}}, \bibinfo {author} {\bibfnamefont {W.}~\bibnamefont {Wang}},\ and\
  \bibinfo {author} {\bibfnamefont {M.~Z.}\ \bibnamefont {Hasan}},\ }\bibfield
  {title} {\bibinfo {title} {Giant and anisotropic many-body spin\textendash
  orbit tunability in a strongly correlated kagome magnet},\ }\href
  {https://doi.org/10.1038/s41586-018-0502-7} {\bibfield  {journal} {\bibinfo
  {journal} {Nature}\ }\textbf {\bibinfo {volume} {562}},\ \bibinfo {pages}
  {91} (\bibinfo {year} {2018})}\BibitemShut {NoStop}%
\bibitem [{\citenamefont {Yin}\ \emph {et~al.}(2019)\citenamefont {Yin},
  \citenamefont {Zhang}, \citenamefont {Chang}, \citenamefont {Wang},
  \citenamefont {Tsirkin}, \citenamefont {Guguchia}, \citenamefont {Lian},
  \citenamefont {Zhou}, \citenamefont {Jiang}, \citenamefont {Belopolski},
  \citenamefont {Shumiya}, \citenamefont {Multer}, \citenamefont {Litskevich},
  \citenamefont {Cochran}, \citenamefont {Lin}, \citenamefont {Wang},
  \citenamefont {Neupert}, \citenamefont {Jia}, \citenamefont {Lei},\ and\
  \citenamefont {Hasan}}]{negativeflatband_hasan}%
  \BibitemOpen
  \bibfield  {author} {\bibinfo {author} {\bibfnamefont {J.-X.}\ \bibnamefont
  {Yin}}, \bibinfo {author} {\bibfnamefont {S.~S.}\ \bibnamefont {Zhang}},
  \bibinfo {author} {\bibfnamefont {G.}~\bibnamefont {Chang}}, \bibinfo
  {author} {\bibfnamefont {Q.}~\bibnamefont {Wang}}, \bibinfo {author}
  {\bibfnamefont {S.~S.}\ \bibnamefont {Tsirkin}}, \bibinfo {author}
  {\bibfnamefont {Z.}~\bibnamefont {Guguchia}}, \bibinfo {author}
  {\bibfnamefont {B.}~\bibnamefont {Lian}}, \bibinfo {author} {\bibfnamefont
  {H.}~\bibnamefont {Zhou}}, \bibinfo {author} {\bibfnamefont {K.}~\bibnamefont
  {Jiang}}, \bibinfo {author} {\bibfnamefont {I.}~\bibnamefont {Belopolski}},
  \bibinfo {author} {\bibfnamefont {N.}~\bibnamefont {Shumiya}}, \bibinfo
  {author} {\bibfnamefont {D.}~\bibnamefont {Multer}}, \bibinfo {author}
  {\bibfnamefont {M.}~\bibnamefont {Litskevich}}, \bibinfo {author}
  {\bibfnamefont {T.~A.}\ \bibnamefont {Cochran}}, \bibinfo {author}
  {\bibfnamefont {H.}~\bibnamefont {Lin}}, \bibinfo {author} {\bibfnamefont
  {Z.}~\bibnamefont {Wang}}, \bibinfo {author} {\bibfnamefont {T.}~\bibnamefont
  {Neupert}}, \bibinfo {author} {\bibfnamefont {S.}~\bibnamefont {Jia}},
  \bibinfo {author} {\bibfnamefont {H.}~\bibnamefont {Lei}},\ and\ \bibinfo
  {author} {\bibfnamefont {M.~Z.}\ \bibnamefont {Hasan}},\ }\bibfield  {title}
  {\bibinfo {title} {Negative flat band magnetism in a spin\textendash
  orbit-coupled correlated kagome magnet},\ }\href
  {https://doi.org/10.1038/s41567-019-0426-7} {\bibfield  {journal} {\bibinfo
  {journal} {Nature Physics}\ }\textbf {\bibinfo {volume} {15}},\ \bibinfo
  {pages} {443} (\bibinfo {year} {2019})}\BibitemShut {NoStop}%
\bibitem [{\citenamefont {Ghimire}\ \emph {et~al.}(2020)\citenamefont
  {Ghimire}, \citenamefont {Dally}, \citenamefont {Poudel}, \citenamefont
  {Jones}, \citenamefont {Michel}, \citenamefont {Magar}, \citenamefont
  {Bleuel}, \citenamefont {McGuire}, \citenamefont {Jiang}, \citenamefont
  {Mitchell}, \citenamefont {Lynn},\ and\ \citenamefont {Mazin}}]{Ghimire2020}%
  \BibitemOpen
  \bibfield  {author} {\bibinfo {author} {\bibfnamefont {N.~J.}\ \bibnamefont
  {Ghimire}}, \bibinfo {author} {\bibfnamefont {R.~L.}\ \bibnamefont {Dally}},
  \bibinfo {author} {\bibfnamefont {L.}~\bibnamefont {Poudel}}, \bibinfo
  {author} {\bibfnamefont {D.~C.}\ \bibnamefont {Jones}}, \bibinfo {author}
  {\bibfnamefont {D.}~\bibnamefont {Michel}}, \bibinfo {author} {\bibfnamefont
  {N.~T.}\ \bibnamefont {Magar}}, \bibinfo {author} {\bibfnamefont
  {M.}~\bibnamefont {Bleuel}}, \bibinfo {author} {\bibfnamefont {M.~A.}\
  \bibnamefont {McGuire}}, \bibinfo {author} {\bibfnamefont {J.~S.}\
  \bibnamefont {Jiang}}, \bibinfo {author} {\bibfnamefont {J.~F.}\ \bibnamefont
  {Mitchell}}, \bibinfo {author} {\bibfnamefont {J.~W.}\ \bibnamefont {Lynn}},\
  and\ \bibinfo {author} {\bibfnamefont {I.~I.}\ \bibnamefont {Mazin}},\
  }\bibfield  {title} {\bibinfo {title} {Competing magnetic phases and
  fluctuation-driven scalar spin chirality in the kagome metal \ch{YMn6Sn6}},\
  }\href {https://doi.org/10.1126/sciadv.abe2680} {\bibfield  {journal}
  {\bibinfo  {journal} {Science Advances}\ }\textbf {\bibinfo {volume} {6}},\
  \bibinfo {pages} {eabe2680} (\bibinfo {year} {2020})},\ \Eprint
  {https://arxiv.org/abs/https://www.science.org/doi/pdf/10.1126/sciadv.abe2680}
  {https://www.science.org/doi/pdf/10.1126/sciadv.abe2680} \BibitemShut
  {NoStop}%
\bibitem [{\citenamefont {Ortiz}\ \emph {et~al.}(2019)\citenamefont {Ortiz},
  \citenamefont {Gomes}, \citenamefont {Morey}, \citenamefont {Winiarski},
  \citenamefont {Bordelon}, \citenamefont {Mangum}, \citenamefont {Oswald},
  \citenamefont {Rodriguez-Rivera}, \citenamefont {Neilson}, \citenamefont
  {Wilson}, \citenamefont {Ertekin}, \citenamefont {McQueen},\ and\
  \citenamefont {Toberer}}]{Ortiz2019}%
  \BibitemOpen
  \bibfield  {author} {\bibinfo {author} {\bibfnamefont {B.~R.}\ \bibnamefont
  {Ortiz}}, \bibinfo {author} {\bibfnamefont {L.~C.}\ \bibnamefont {Gomes}},
  \bibinfo {author} {\bibfnamefont {J.~R.}\ \bibnamefont {Morey}}, \bibinfo
  {author} {\bibfnamefont {M.}~\bibnamefont {Winiarski}}, \bibinfo {author}
  {\bibfnamefont {M.}~\bibnamefont {Bordelon}}, \bibinfo {author}
  {\bibfnamefont {J.~S.}\ \bibnamefont {Mangum}}, \bibinfo {author}
  {\bibfnamefont {I.~W.~H.}\ \bibnamefont {Oswald}}, \bibinfo {author}
  {\bibfnamefont {J.~A.}\ \bibnamefont {Rodriguez-Rivera}}, \bibinfo {author}
  {\bibfnamefont {J.~R.}\ \bibnamefont {Neilson}}, \bibinfo {author}
  {\bibfnamefont {S.~D.}\ \bibnamefont {Wilson}}, \bibinfo {author}
  {\bibfnamefont {E.}~\bibnamefont {Ertekin}}, \bibinfo {author} {\bibfnamefont
  {T.~M.}\ \bibnamefont {McQueen}},\ and\ \bibinfo {author} {\bibfnamefont
  {E.~S.}\ \bibnamefont {Toberer}},\ }\bibfield  {title} {\bibinfo {title} {New
  kagome prototype materials: discovery of \ch{KV3Sb5}, \ch{RbV3Sb5}, and
  \ch{CsV3Sb5}},\ }\href {https://doi.org/10.1103/PhysRevMaterials.3.094407}
  {\bibfield  {journal} {\bibinfo  {journal} {Phys. Rev. Materials}\ }\textbf
  {\bibinfo {volume} {3}},\ \bibinfo {pages} {094407} (\bibinfo {year}
  {2019})}\BibitemShut {NoStop}%
\bibitem [{\citenamefont {Ortiz}\ \emph {et~al.}(2020)\citenamefont {Ortiz},
  \citenamefont {Teicher}, \citenamefont {Hu}, \citenamefont {Zuo},
  \citenamefont {Sarte}, \citenamefont {Schueller}, \citenamefont {Abeykoon},
  \citenamefont {Krogstad}, \citenamefont {Rosenkranz}, \citenamefont {Osborn},
  \citenamefont {Seshadri}, \citenamefont {Balents}, \citenamefont {He},\ and\
  \citenamefont {Wilson}}]{Ortiz2020}%
  \BibitemOpen
  \bibfield  {author} {\bibinfo {author} {\bibfnamefont {B.~R.}\ \bibnamefont
  {Ortiz}}, \bibinfo {author} {\bibfnamefont {S.~M.~L.}\ \bibnamefont
  {Teicher}}, \bibinfo {author} {\bibfnamefont {Y.}~\bibnamefont {Hu}},
  \bibinfo {author} {\bibfnamefont {J.~L.}\ \bibnamefont {Zuo}}, \bibinfo
  {author} {\bibfnamefont {P.~M.}\ \bibnamefont {Sarte}}, \bibinfo {author}
  {\bibfnamefont {E.~C.}\ \bibnamefont {Schueller}}, \bibinfo {author}
  {\bibfnamefont {A.~M.~M.}\ \bibnamefont {Abeykoon}}, \bibinfo {author}
  {\bibfnamefont {M.~J.}\ \bibnamefont {Krogstad}}, \bibinfo {author}
  {\bibfnamefont {S.}~\bibnamefont {Rosenkranz}}, \bibinfo {author}
  {\bibfnamefont {R.}~\bibnamefont {Osborn}}, \bibinfo {author} {\bibfnamefont
  {R.}~\bibnamefont {Seshadri}}, \bibinfo {author} {\bibfnamefont
  {L.}~\bibnamefont {Balents}}, \bibinfo {author} {\bibfnamefont
  {J.}~\bibnamefont {He}},\ and\ \bibinfo {author} {\bibfnamefont {S.~D.}\
  \bibnamefont {Wilson}},\ }\bibfield  {title} {\bibinfo {title} {\ch{CsV3Sb5}:
  A {{Z$_2$}} topological kagome metal with a superconducting ground state},\
  }\href {https://doi.org/10.1103/PhysRevLett.125.247002} {\bibfield  {journal}
  {\bibinfo  {journal} {Phys. Rev. Lett.}\ }\textbf {\bibinfo {volume} {125}},\
  \bibinfo {pages} {247002} (\bibinfo {year} {2020})}\BibitemShut {NoStop}%
\bibitem [{\citenamefont {Yin}\ \emph {et~al.}(2020)\citenamefont {Yin},
  \citenamefont {Ma}, \citenamefont {Cochran}, \citenamefont {Xu},
  \citenamefont {Zhang}, \citenamefont {Tien}, \citenamefont {Shumiya},
  \citenamefont {Cheng}, \citenamefont {Jiang}, \citenamefont {Lian},
  \citenamefont {Song}, \citenamefont {Chang}, \citenamefont {Belopolski},
  \citenamefont {Multer}, \citenamefont {Litskevich}, \citenamefont {Cheng},
  \citenamefont {Yang}, \citenamefont {Swidler}, \citenamefont {Zhou},
  \citenamefont {Lin}, \citenamefont {Neupert}, \citenamefont {Wang},
  \citenamefont {Yao}, \citenamefont {Chang}, \citenamefont {Jia},\ and\
  \citenamefont {Zahid~Hasan}}]{TMS_hasan}%
  \BibitemOpen
  \bibfield  {author} {\bibinfo {author} {\bibfnamefont {J.-X.}\ \bibnamefont
  {Yin}}, \bibinfo {author} {\bibfnamefont {W.}~\bibnamefont {Ma}}, \bibinfo
  {author} {\bibfnamefont {T.~A.}\ \bibnamefont {Cochran}}, \bibinfo {author}
  {\bibfnamefont {X.}~\bibnamefont {Xu}}, \bibinfo {author} {\bibfnamefont
  {S.~S.}\ \bibnamefont {Zhang}}, \bibinfo {author} {\bibfnamefont {H.-J.}\
  \bibnamefont {Tien}}, \bibinfo {author} {\bibfnamefont {N.}~\bibnamefont
  {Shumiya}}, \bibinfo {author} {\bibfnamefont {G.}~\bibnamefont {Cheng}},
  \bibinfo {author} {\bibfnamefont {K.}~\bibnamefont {Jiang}}, \bibinfo
  {author} {\bibfnamefont {B.}~\bibnamefont {Lian}}, \bibinfo {author}
  {\bibfnamefont {Z.}~\bibnamefont {Song}}, \bibinfo {author} {\bibfnamefont
  {G.}~\bibnamefont {Chang}}, \bibinfo {author} {\bibfnamefont
  {I.}~\bibnamefont {Belopolski}}, \bibinfo {author} {\bibfnamefont
  {D.}~\bibnamefont {Multer}}, \bibinfo {author} {\bibfnamefont
  {M.}~\bibnamefont {Litskevich}}, \bibinfo {author} {\bibfnamefont {Z.-J.}\
  \bibnamefont {Cheng}}, \bibinfo {author} {\bibfnamefont {X.~P.}\ \bibnamefont
  {Yang}}, \bibinfo {author} {\bibfnamefont {B.}~\bibnamefont {Swidler}},
  \bibinfo {author} {\bibfnamefont {H.}~\bibnamefont {Zhou}}, \bibinfo {author}
  {\bibfnamefont {H.}~\bibnamefont {Lin}}, \bibinfo {author} {\bibfnamefont
  {T.}~\bibnamefont {Neupert}}, \bibinfo {author} {\bibfnamefont
  {Z.}~\bibnamefont {Wang}}, \bibinfo {author} {\bibfnamefont {N.}~\bibnamefont
  {Yao}}, \bibinfo {author} {\bibfnamefont {T.-R.}\ \bibnamefont {Chang}},
  \bibinfo {author} {\bibfnamefont {S.}~\bibnamefont {Jia}},\ and\ \bibinfo
  {author} {\bibfnamefont {M.}~\bibnamefont {Zahid~Hasan}},\ }\bibfield
  {title} {\bibinfo {title} {Quantum-limit {{Chern}} topological magnetism in
  \ch{TbMn6Sn6}},\ }\href {https://doi.org/10.1038/s41586-020-2482-7}
  {\bibfield  {journal} {\bibinfo  {journal} {Nature}\ }\textbf {\bibinfo
  {volume} {583}},\ \bibinfo {pages} {533} (\bibinfo {year}
  {2020})}\BibitemShut {NoStop}%
\bibitem [{\citenamefont {Ma}\ \emph {et~al.}(2021)\citenamefont {Ma},
  \citenamefont {Xu}, \citenamefont {Yin}, \citenamefont {Yang}, \citenamefont
  {Zhou}, \citenamefont {Cheng}, \citenamefont {Huang}, \citenamefont {Qu},
  \citenamefont {Wang}, \citenamefont {Hasan},\ and\ \citenamefont
  {Jia}}]{RMS_engineering_Jia}%
  \BibitemOpen
  \bibfield  {author} {\bibinfo {author} {\bibfnamefont {W.}~\bibnamefont
  {Ma}}, \bibinfo {author} {\bibfnamefont {X.}~\bibnamefont {Xu}}, \bibinfo
  {author} {\bibfnamefont {J.-X.}\ \bibnamefont {Yin}}, \bibinfo {author}
  {\bibfnamefont {H.}~\bibnamefont {Yang}}, \bibinfo {author} {\bibfnamefont
  {H.}~\bibnamefont {Zhou}}, \bibinfo {author} {\bibfnamefont {Z.-J.}\
  \bibnamefont {Cheng}}, \bibinfo {author} {\bibfnamefont {Y.}~\bibnamefont
  {Huang}}, \bibinfo {author} {\bibfnamefont {Z.}~\bibnamefont {Qu}}, \bibinfo
  {author} {\bibfnamefont {F.}~\bibnamefont {Wang}}, \bibinfo {author}
  {\bibfnamefont {M.~Z.}\ \bibnamefont {Hasan}},\ and\ \bibinfo {author}
  {\bibfnamefont {S.}~\bibnamefont {Jia}},\ }\bibfield  {title} {\bibinfo
  {title} {Rare {{Earth Engineering}} in {{RMn$_6$Sn$_6$}} ( {{R}} = {{Gd}} -
  {{Tm}} , {{Lu}}) {{Topological Kagome Magnets}}},\ }\href
  {https://doi.org/10.1103/PhysRevLett.126.246602} {\bibfield  {journal}
  {\bibinfo  {journal} {Physical Review Letters}\ }\textbf {\bibinfo {volume}
  {126}},\ \bibinfo {pages} {246602} (\bibinfo {year} {2021})}\BibitemShut
  {NoStop}%
\bibitem [{\citenamefont {Chen}\ \emph {et~al.}(2021)\citenamefont {Chen},
  \citenamefont {Le}, \citenamefont {Fu}, \citenamefont {Lin}, \citenamefont
  {Schnelle}, \citenamefont {Sun},\ and\ \citenamefont
  {Felser}}]{LMS_anomhall-FM_felser}%
  \BibitemOpen
  \bibfield  {author} {\bibinfo {author} {\bibfnamefont {D.}~\bibnamefont
  {Chen}}, \bibinfo {author} {\bibfnamefont {C.}~\bibnamefont {Le}}, \bibinfo
  {author} {\bibfnamefont {C.}~\bibnamefont {Fu}}, \bibinfo {author}
  {\bibfnamefont {H.}~\bibnamefont {Lin}}, \bibinfo {author} {\bibfnamefont
  {W.}~\bibnamefont {Schnelle}}, \bibinfo {author} {\bibfnamefont
  {Y.}~\bibnamefont {Sun}},\ and\ \bibinfo {author} {\bibfnamefont
  {C.}~\bibnamefont {Felser}},\ }\bibfield  {title} {\bibinfo {title} {Large
  anomalous {{Hall}} effect in the {{Kagome Ferromagnet}} \ch{LiMn6Sn6}},\
  }\href {https://doi.org/10.1103/PhysRevB.103.144410} {\bibfield  {journal}
  {\bibinfo  {journal} {Physical Review B}\ }\textbf {\bibinfo {volume}
  {103}},\ \bibinfo {pages} {144410} (\bibinfo {year} {2021})}\BibitemShut
  {NoStop}%
\bibitem [{\citenamefont {Pokharel}\ \emph {et~al.}(2021)\citenamefont
  {Pokharel}, \citenamefont {Teicher}, \citenamefont {Ortiz}, \citenamefont
  {Sarte}, \citenamefont {Wu}, \citenamefont {Peng}, \citenamefont {He},
  \citenamefont {Seshadri},\ and\ \citenamefont {Wilson}}]{YVSGVS_wilson}%
  \BibitemOpen
  \bibfield  {author} {\bibinfo {author} {\bibfnamefont {G.}~\bibnamefont
  {Pokharel}}, \bibinfo {author} {\bibfnamefont {S.~M.~L.}\ \bibnamefont
  {Teicher}}, \bibinfo {author} {\bibfnamefont {B.~R.}\ \bibnamefont {Ortiz}},
  \bibinfo {author} {\bibfnamefont {P.~M.}\ \bibnamefont {Sarte}}, \bibinfo
  {author} {\bibfnamefont {G.}~\bibnamefont {Wu}}, \bibinfo {author}
  {\bibfnamefont {S.}~\bibnamefont {Peng}}, \bibinfo {author} {\bibfnamefont
  {J.}~\bibnamefont {He}}, \bibinfo {author} {\bibfnamefont {R.}~\bibnamefont
  {Seshadri}},\ and\ \bibinfo {author} {\bibfnamefont {S.~D.}\ \bibnamefont
  {Wilson}},\ }\bibfield  {title} {\bibinfo {title} {Study of the electronic
  properties of topological kagome metals \ch{YV6Sn6} and \ch{GdV6Sn6}},\
  }\href@noop {} {\bibfield  {journal} {\bibinfo  {journal} {arXiv:2109.07394
  [cond-mat]}\ } (\bibinfo {year} {2021})},\ \Eprint
  {https://arxiv.org/abs/2109.07394} {arXiv:2109.07394 [cond-mat]} \BibitemShut
  {NoStop}%
\bibitem [{\citenamefont {Ishikawa}\ \emph {et~al.}(2021)\citenamefont
  {Ishikawa}, \citenamefont {Yajima}, \citenamefont {Kawamura}, \citenamefont
  {Mitamura},\ and\ \citenamefont {Kindo}}]{Ishikawa2021}%
  \BibitemOpen
  \bibfield  {author} {\bibinfo {author} {\bibfnamefont {H.}~\bibnamefont
  {Ishikawa}}, \bibinfo {author} {\bibfnamefont {T.}~\bibnamefont {Yajima}},
  \bibinfo {author} {\bibfnamefont {M.}~\bibnamefont {Kawamura}}, \bibinfo
  {author} {\bibfnamefont {H.}~\bibnamefont {Mitamura}},\ and\ \bibinfo
  {author} {\bibfnamefont {K.}~\bibnamefont {Kindo}},\ }\bibfield  {title}
  {\bibinfo {title} {Gdv6sn6: A multi-carrier metal with non-magnetic
  3d-electron kagome bands and 4f-electron magnetism},\ }\href
  {https://doi.org/10.7566/JPSJ.90.124704} {\bibfield  {journal} {\bibinfo
  {journal} {Journal of the Physical Society of Japan}\ }\textbf {\bibinfo
  {volume} {90}},\ \bibinfo {pages} {124704} (\bibinfo {year} {2021})},\
  \Eprint {https://arxiv.org/abs/https://doi.org/10.7566/JPSJ.90.124704}
  {https://doi.org/10.7566/JPSJ.90.124704} \BibitemShut {NoStop}%
\bibitem [{\citenamefont {{Suriya Arachchige}}\ \emph
  {et~al.}(2022)\citenamefont {{Suriya Arachchige}}, \citenamefont {{Meier}},
  \citenamefont {{Marshall}}, \citenamefont {{Matsuoka}}, \citenamefont
  {{Xue}}, \citenamefont {{McGuire}}, \citenamefont {{Hermann}}, \citenamefont
  {{Cao}},\ and\ \citenamefont {{Mandrus}}}]{Arachchige2022}%
  \BibitemOpen
  \bibfield  {author} {\bibinfo {author} {\bibfnamefont {H.~W.}\ \bibnamefont
  {{Suriya Arachchige}}}, \bibinfo {author} {\bibfnamefont {W.~R.}\
  \bibnamefont {{Meier}}}, \bibinfo {author} {\bibfnamefont {M.}~\bibnamefont
  {{Marshall}}}, \bibinfo {author} {\bibfnamefont {T.}~\bibnamefont
  {{Matsuoka}}}, \bibinfo {author} {\bibfnamefont {R.}~\bibnamefont {{Xue}}},
  \bibinfo {author} {\bibfnamefont {M.~A.}\ \bibnamefont {{McGuire}}}, \bibinfo
  {author} {\bibfnamefont {R.~P.}\ \bibnamefont {{Hermann}}}, \bibinfo {author}
  {\bibfnamefont {H.}~\bibnamefont {{Cao}}},\ and\ \bibinfo {author}
  {\bibfnamefont {D.}~\bibnamefont {{Mandrus}}},\ }\bibfield  {title} {\bibinfo
  {title} {{Charge density wave in kagome lattice intermetallic
  \ch{ScV6Sn6}}},\ }\href@noop {} {\bibfield  {journal} {\bibinfo  {journal}
  {arXiv e-prints}\ ,\ \bibinfo {eid} {arXiv:2205.04582}} (\bibinfo {year}
  {2022})},\ \Eprint {https://arxiv.org/abs/2205.04582} {arXiv:2205.04582
  [cond-mat.str-el]} \BibitemShut {NoStop}%
\bibitem [{\citenamefont {Peng}\ \emph {et~al.}(2021)\citenamefont {Peng},
  \citenamefont {Han}, \citenamefont {Pokharel}, \citenamefont {Shen},
  \citenamefont {Li}, \citenamefont {Hashimoto}, \citenamefont {Lu},
  \citenamefont {Ortiz}, \citenamefont {Luo}, \citenamefont {Li}, \citenamefont
  {Guo}, \citenamefont {Wang}, \citenamefont {Cui}, \citenamefont {Sun},
  \citenamefont {Qiao}, \citenamefont {Wilson},\ and\ \citenamefont
  {He}}]{GVSHVS_arpes_He}%
  \BibitemOpen
  \bibfield  {author} {\bibinfo {author} {\bibfnamefont {S.}~\bibnamefont
  {Peng}}, \bibinfo {author} {\bibfnamefont {Y.}~\bibnamefont {Han}}, \bibinfo
  {author} {\bibfnamefont {G.}~\bibnamefont {Pokharel}}, \bibinfo {author}
  {\bibfnamefont {J.}~\bibnamefont {Shen}}, \bibinfo {author} {\bibfnamefont
  {Z.}~\bibnamefont {Li}}, \bibinfo {author} {\bibfnamefont {M.}~\bibnamefont
  {Hashimoto}}, \bibinfo {author} {\bibfnamefont {D.}~\bibnamefont {Lu}},
  \bibinfo {author} {\bibfnamefont {B.~R.}\ \bibnamefont {Ortiz}}, \bibinfo
  {author} {\bibfnamefont {Y.}~\bibnamefont {Luo}}, \bibinfo {author}
  {\bibfnamefont {H.}~\bibnamefont {Li}}, \bibinfo {author} {\bibfnamefont
  {M.}~\bibnamefont {Guo}}, \bibinfo {author} {\bibfnamefont {B.}~\bibnamefont
  {Wang}}, \bibinfo {author} {\bibfnamefont {S.}~\bibnamefont {Cui}}, \bibinfo
  {author} {\bibfnamefont {Z.}~\bibnamefont {Sun}}, \bibinfo {author}
  {\bibfnamefont {Z.}~\bibnamefont {Qiao}}, \bibinfo {author} {\bibfnamefont
  {S.~D.}\ \bibnamefont {Wilson}},\ and\ \bibinfo {author} {\bibfnamefont
  {J.}~\bibnamefont {He}},\ }\bibfield  {title} {\bibinfo {title} {Realizing
  {{Kagome Band Structure}} in {{Two}}-{{Dimensional Kagome Surface States}} of
  {{RV$_6$Sn$_6$}} ({{R}} = {{Gd}} , {{Ho}})},\ }\href
  {https://doi.org/10.1103/PhysRevLett.127.266401} {\bibfield  {journal}
  {\bibinfo  {journal} {Physical Review Letters}\ }\textbf {\bibinfo {volume}
  {127}},\ \bibinfo {pages} {266401} (\bibinfo {year} {2021})}\BibitemShut
  {NoStop}%
\bibitem [{\citenamefont {Gao}\ \emph {et~al.}(2021)\citenamefont {Gao},
  \citenamefont {Shen}, \citenamefont {Wang}, \citenamefont {Shi},
  \citenamefont {Zhao}, \citenamefont {Li}, \citenamefont {Cao}, \citenamefont
  {Pei}, \citenamefont {Ge}, \citenamefont {Li}, \citenamefont {Li},
  \citenamefont {Chen}, \citenamefont {Yan},\ and\ \citenamefont
  {Qi}}]{RMS_anomhall-FIM_qi}%
  \BibitemOpen
  \bibfield  {author} {\bibinfo {author} {\bibfnamefont {L.}~\bibnamefont
  {Gao}}, \bibinfo {author} {\bibfnamefont {S.}~\bibnamefont {Shen}}, \bibinfo
  {author} {\bibfnamefont {Q.}~\bibnamefont {Wang}}, \bibinfo {author}
  {\bibfnamefont {W.}~\bibnamefont {Shi}}, \bibinfo {author} {\bibfnamefont
  {Y.}~\bibnamefont {Zhao}}, \bibinfo {author} {\bibfnamefont {C.}~\bibnamefont
  {Li}}, \bibinfo {author} {\bibfnamefont {W.}~\bibnamefont {Cao}}, \bibinfo
  {author} {\bibfnamefont {C.}~\bibnamefont {Pei}}, \bibinfo {author}
  {\bibfnamefont {J.-Y.}\ \bibnamefont {Ge}}, \bibinfo {author} {\bibfnamefont
  {G.}~\bibnamefont {Li}}, \bibinfo {author} {\bibfnamefont {J.}~\bibnamefont
  {Li}}, \bibinfo {author} {\bibfnamefont {Y.}~\bibnamefont {Chen}}, \bibinfo
  {author} {\bibfnamefont {S.}~\bibnamefont {Yan}},\ and\ \bibinfo {author}
  {\bibfnamefont {Y.}~\bibnamefont {Qi}},\ }\bibfield  {title} {\bibinfo
  {title} {Anomalous {{Hall}} effect in ferrimagnetic metal {{RMn$_6$Sn$_6$}}
  ({{R}} = {{Tb}}, {{Dy}}, {{Ho}}) with clean {{Mn}} kagome lattice},\ }\href
  {https://doi.org/10.1063/5.0061260} {\bibfield  {journal} {\bibinfo
  {journal} {Applied Physics Letters}\ }\textbf {\bibinfo {volume} {119}},\
  \bibinfo {pages} {092405} (\bibinfo {year} {2021})}\BibitemShut {NoStop}%
\bibitem [{\citenamefont {Clatterbuck}\ and\ \citenamefont
  {Jr}(1999)}]{RMS_many_gschneidner}%
  \BibitemOpen
  \bibfield  {author} {\bibinfo {author} {\bibfnamefont {D.~M.}\ \bibnamefont
  {Clatterbuck}}\ and\ \bibinfo {author} {\bibfnamefont {K.~A.~G.}\
  \bibnamefont {Jr}},\ }\bibfield  {title} {\bibinfo {title} {Magnetic
  properties of {{RMn$_6$Sn$_6$ (R = Tb, Ho, Er, Tm, Lu)}} single crystals},\
  }\href {https://doi.org/https://doi.org/10.1016/S0304-8853(99)00571-5}
  {\bibfield  {journal} {\bibinfo  {journal} {Journal of Magnetism and Magnetic
  Materials}\ ,\ \bibinfo {pages} {17}} (\bibinfo {year} {1999})}\BibitemShut
  {NoStop}%
\bibitem [{\citenamefont {Canfield}\ \emph {et~al.}(2016)\citenamefont
  {Canfield}, \citenamefont {Kong}, \citenamefont {Kaluarachchi},\ and\
  \citenamefont {Jo}}]{Canfield2016}%
  \BibitemOpen
  \bibfield  {author} {\bibinfo {author} {\bibfnamefont {P.~C.}\ \bibnamefont
  {Canfield}}, \bibinfo {author} {\bibfnamefont {T.}~\bibnamefont {Kong}},
  \bibinfo {author} {\bibfnamefont {U.~S.}\ \bibnamefont {Kaluarachchi}},\ and\
  \bibinfo {author} {\bibfnamefont {N.~H.}\ \bibnamefont {Jo}},\ }\bibfield
  {title} {\bibinfo {title} {Use of frit-disc crucibles for routine and
  exploratory solution growth of single crystalline samples},\ }\href
  {https://doi.org/10.1080/14786435.2015.1122248} {\bibfield  {journal}
  {\bibinfo  {journal} {Philosophical Magazine}\ }\textbf {\bibinfo {volume}
  {96}},\ \bibinfo {pages} {84} (\bibinfo {year} {2016})},\ \Eprint
  {https://arxiv.org/abs/https://doi.org/10.1080/14786435.2015.1122248}
  {https://doi.org/10.1080/14786435.2015.1122248} \BibitemShut {NoStop}%
\bibitem [{\citenamefont {Lutterotti}\ \emph {et~al.}(2007)\citenamefont
  {Lutterotti}, \citenamefont {Bortolotti}, \citenamefont {Ischia},\ and\
  \citenamefont {Lonardelli}}]{MAUD}%
  \BibitemOpen
  \bibfield  {author} {\bibinfo {author} {\bibfnamefont {L.}~\bibnamefont
  {Lutterotti}}, \bibinfo {author} {\bibfnamefont {M.}~\bibnamefont
  {Bortolotti}}, \bibinfo {author} {\bibfnamefont {G.}~\bibnamefont {Ischia}},\
  and\ \bibinfo {author} {\bibfnamefont {I.}~\bibnamefont {Lonardelli}},\
  }\bibfield  {title} {\bibinfo {title} {Rietveld texture analysis from
  diffraction images},\ }\href@noop {} {\bibfield  {journal} {\bibinfo
  {journal} {European Powder Diffraction Conference}\ }\textbf {\bibinfo
  {volume} {26}},\ \bibinfo {pages} {125} (\bibinfo {year} {2007})}\BibitemShut
  {NoStop}%
\bibitem [{\citenamefont {Ke}(2019)}]{ke2019prb}%
  \BibitemOpen
  \bibfield  {author} {\bibinfo {author} {\bibfnamefont {L.}~\bibnamefont
  {Ke}},\ }\bibfield  {title} {\bibinfo {title} {Intersublattice
  magnetocrystalline anisotropy using a realistic tight-binding method based on
  maximally localized {Wannier} functions},\ }\href
  {https://doi.org/10.1103/PhysRevB.99.054418} {\bibfield  {journal} {\bibinfo
  {journal} {Phys. Rev. B}\ }\textbf {\bibinfo {volume} {99}},\ \bibinfo
  {pages} {054418} (\bibinfo {year} {2019})}\BibitemShut {NoStop}%
\bibitem [{\citenamefont {Lee}\ \emph {et~al.}(2022)\citenamefont {Lee},
  \citenamefont {Skomski}, \citenamefont {Wang}, \citenamefont {Orth},
  \citenamefont {Pathak}, \citenamefont {Harmon}, \citenamefont {McQueeney},
  \citenamefont {Mazin},\ and\ \citenamefont {Ke}}]{lee2022arxiv}%
  \BibitemOpen
  \bibfield  {author} {\bibinfo {author} {\bibfnamefont {Y.}~\bibnamefont
  {Lee}}, \bibinfo {author} {\bibfnamefont {R.}~\bibnamefont {Skomski}},
  \bibinfo {author} {\bibfnamefont {X.}~\bibnamefont {Wang}}, \bibinfo {author}
  {\bibfnamefont {P.~P.}\ \bibnamefont {Orth}}, \bibinfo {author}
  {\bibfnamefont {A.~K.}\ \bibnamefont {Pathak}}, \bibinfo {author}
  {\bibfnamefont {B.~N.}\ \bibnamefont {Harmon}}, \bibinfo {author}
  {\bibfnamefont {R.~J.}\ \bibnamefont {McQueeney}}, \bibinfo {author}
  {\bibfnamefont {I.~I.}\ \bibnamefont {Mazin}},\ and\ \bibinfo {author}
  {\bibfnamefont {L.}~\bibnamefont {Ke}},\ }\bibfield  {title} {\bibinfo
  {title} {{Interplay between magnetism and band topology in Kagome magnets
  $R$Mn$_6$Sn$_6$}},\ }\href@noop {} {\bibfield  {journal} {\bibinfo  {journal}
  {arXiv}\ }\textbf {\bibinfo {volume} {2201}},\ \bibinfo {pages} {11265}
  (\bibinfo {year} {2022})},\ \Eprint
  {https://arxiv.org/abs/https://arxiv.org/abs/2201.11265}
  {https://arxiv.org/abs/2201.11265} \BibitemShut {NoStop}%
\bibitem [{\citenamefont {Beauvillain}\ \emph {et~al.}(1978)\citenamefont
  {Beauvillain}, \citenamefont {Renard}, \citenamefont {Laursen},\ and\
  \citenamefont {Walker}}]{LiHoF4_walker}%
  \BibitemOpen
  \bibfield  {author} {\bibinfo {author} {\bibfnamefont {P.}~\bibnamefont
  {Beauvillain}}, \bibinfo {author} {\bibfnamefont {J.~P.}\ \bibnamefont
  {Renard}}, \bibinfo {author} {\bibfnamefont {I.}~\bibnamefont {Laursen}},\
  and\ \bibinfo {author} {\bibfnamefont {P.~J.}\ \bibnamefont {Walker}},\
  }\bibfield  {title} {\bibinfo {title} {Critical behavior of the magnetic
  susceptibility of the uniaxial ferromagnet \ch{LiHoF4}},\ }\href
  {https://doi.org/10.1103/PhysRevB.18.3360} {\bibfield  {journal} {\bibinfo
  {journal} {Physical Review B}\ }\textbf {\bibinfo {volume} {18}},\ \bibinfo
  {pages} {3360} (\bibinfo {year} {1978})}\BibitemShut {NoStop}%
\bibitem [{\citenamefont {Watts~S.M.}\ and\ \citenamefont {von
  Molnar}(2000)}]{Hallpaper}%
  \BibitemOpen
  \bibfield  {author} {\bibinfo {author} {\bibfnamefont {S.}~\bibnamefont
  {Watts~S.M.}, \bibfnamefont {Wirth}}\ and\ \bibinfo {author} {\bibfnamefont
  {S.}~\bibnamefont {von Molnar}},\ }\bibfield  {title} {\bibinfo {title}
  {Evidence for two-band magnetotransport in half-metallic chromium dioxide},\
  }\href {https://doi.org/10.1103/PhysRevB.18.3360} {\bibfield  {journal}
  {\bibinfo  {journal} {Physical Review B}\ }\textbf {\bibinfo {volume} {61}},\
  \bibinfo {pages} {9621} (\bibinfo {year} {2000})}\BibitemShut {NoStop}%
\bibitem [{\citenamefont {Morin}\ and\ \citenamefont
  {Schmitt}(1988)}]{quadrupoleinteractions_morin}%
  \BibitemOpen
  \bibfield  {author} {\bibinfo {author} {\bibfnamefont {P.}~\bibnamefont
  {Morin}}\ and\ \bibinfo {author} {\bibfnamefont {D.}~\bibnamefont
  {Schmitt}},\ }\bibfield  {title} {\bibinfo {title} {{{Q}}uadrupole
  interactions in rare-earth intermetallic compounds},\ }\href
  {https://doi.org/10.1051/jphyscol:19888144} {\bibfield  {journal} {\bibinfo
  {journal} {Le Journal de Physique Colloques}\ }\textbf {\bibinfo {volume}
  {49}},\ \bibinfo {pages} {C8} (\bibinfo {year} {1988})}\BibitemShut {NoStop}%
\bibitem [{\citenamefont {Maharaj}\ \emph {et~al.}(2017)\citenamefont
  {Maharaj}, \citenamefont {Rosenberg}, \citenamefont {Hristov}, \citenamefont
  {Berg}, \citenamefont {Fernandes}, \citenamefont {Fisher},\ and\
  \citenamefont {Kivelson}}]{transising_maharaj}%
  \BibitemOpen
  \bibfield  {author} {\bibinfo {author} {\bibfnamefont {A.~V.}\ \bibnamefont
  {Maharaj}}, \bibinfo {author} {\bibfnamefont {E.~W.}\ \bibnamefont
  {Rosenberg}}, \bibinfo {author} {\bibfnamefont {A.~T.}\ \bibnamefont
  {Hristov}}, \bibinfo {author} {\bibfnamefont {E.}~\bibnamefont {Berg}},
  \bibinfo {author} {\bibfnamefont {R.~M.}\ \bibnamefont {Fernandes}}, \bibinfo
  {author} {\bibfnamefont {I.~R.}\ \bibnamefont {Fisher}},\ and\ \bibinfo
  {author} {\bibfnamefont {S.~A.}\ \bibnamefont {Kivelson}},\ }\bibfield
  {title} {\bibinfo {title} {Transverse fields to tune an {{Ising}}-nematic
  quantum phase transition},\ }\href {https://doi.org/10.1073/pnas.1712533114}
  {\bibfield  {journal} {\bibinfo  {journal} {Proceedings of the National
  Academy of Sciences}\ }\textbf {\bibinfo {volume} {114}},\ \bibinfo {pages}
  {13430} (\bibinfo {year} {2017})}\BibitemShut {NoStop}%
\bibitem [{\citenamefont {Tian}\ \emph {et~al.}(2009)\citenamefont {Tian},
  \citenamefont {Ye},\ and\ \citenamefont {Jin}}]{AHEpaper}%
  \BibitemOpen
  \bibfield  {author} {\bibinfo {author} {\bibfnamefont {Y.}~\bibnamefont
  {Tian}}, \bibinfo {author} {\bibfnamefont {L.}~\bibnamefont {Ye}},\ and\
  \bibinfo {author} {\bibfnamefont {X.}~\bibnamefont {Jin}},\ }\bibfield
  {title} {\bibinfo {title} {Proper {{Scaling}} of the {{Anomalous Hall
  Effect}}},\ }\href {https://doi.org/10.1103/PhysRevLett.103.087206}
  {\bibfield  {journal} {\bibinfo  {journal} {Physical Review Letters}\
  }\textbf {\bibinfo {volume} {103}},\ \bibinfo {pages} {087206} (\bibinfo
  {year} {2009})}\BibitemShut {NoStop}%
\end{thebibliography}%

\end{document}